\documentclass[aps,prb,twocolumn]{revtex4} 

\usepackage{graphicx}
\usepackage{dcolumn}
\usepackage{bm}
\usepackage{amsmath}  

\newcommand{\comment}[1]{}

\begin{document}
\renewcommand{\theequation}{\arabic{section}.\arabic{equation}}

\title{Quantum Stochastic Molecular Dynamics Simulations of
the Viscosity of Superfluid Helium}


\author{Phil Attard}
\affiliation{
{\tt phil.attard1@gmail.com}  March 20--June 13 2023}
\noindent {\tt  Projects/QSM23/Paper/QSMD.tex}


\begin{abstract}
Decoherent quantum equations of motion are derived
that yield the trajectory of an open quantum system.
The  viscosity of superfluid Lennard-Jones helium-4 is obtained
with a quantum stochastic molecular dynamics algorithm.
The momentum state occupancy entropy is counted with
a continuous representation of boson number
and averages are obtained with umbrella sampling.
Instantaneous snapshots of the Bose-Einstein condensed system
show multiple highly occupied momentum states.
The viscosity is obtained from the Onsager-Green-Kubo relation
with the time correlation function modified in the quantum case.
On the saturation curve, at higher temperatures
the viscosities of the classical and quantum liquids are equal.
With decreasing temperature
the viscosity of the classical liquid increases
whereas that of the quantum liquid decreases.
Below the $\lambda$-transition the viscosity
lies significantly below the classical value,
being small but positive due to the mixture
of condensed and uncondensed bosons.
The computed trajectories give a physical explanation
of the molecular mechanism for superfluidity.
\end{abstract}


\maketitle

%
\section{Introduction}
\setcounter{equation}{0} \setcounter{subsubsection}{0}
%

This paper explores the nature of Bose-Einstein condensation
and of superfluid viscosity
by using quantum stochastic molecular dynamics (QSMD) computer simulations.
A feature of the calculations is that the bosons interact with each other
via the Lennard-Jones pair potential with helium-4 parameters.
This gives a much more realistic molecular picture
of the Bose-Einstein condensate
than the original ideal boson (ie.\ non-interacting)
calculations of F. London (1938),
which showed that the $\lambda$-transition in $^4$He
was due to Bose-Einstein condensation.
It also gives a more realistic picture of superfluid flow
than the original two-fluid model of Tisza,
which, based on F. London's ideas,
also invokes non-interacting bosons (Tisza 1938, Balibar 2017).

Although today many accept without question
Bose-Einstein condensation as the origin of the $\lambda$-transition
and superfluidity,
in the past this has been seriously questioned.
Landau  objected:
\begin{center}
\parbox{7.5cm}{
[F. London and] `L. Tisza suggested
that helium II should be considered as a degenerate ideal Bose gas\ldots
This point of view, however, cannot be considered as satisfactory\ldots
nothing would prevent atoms in a normal state from colliding with
excited atoms, ie.\ when moving through the liquid they would
experience a friction and there would be no superfluidity at all. In
this way the explanation advanced by Tisza not only has no
foundation in his suggestions but is in direct contradiction to
them.' (Landau 1941)
}\end{center}
One should take Landau's objection seriously,
if for no other reason than that
he was awarded the Nobel prize for this particular paper.
In the conclusion to the present paper a detailed molecular-level rebuttal
of Landau's argument is given.

Feynman was also critical of the two-fluid model of superfluidity:
\begin{center}
\parbox{7.5cm}{
`The division into a normal fluid and a superfluid
although yielding a simple model for understanding the final equations
appears artificial from a microscopic point of view.'
(Feynman 1954)
}\end{center}
The present results agree with this broad characterization by Feynman
in that they do not support the binary division that underlies
the conventional interpretation of Bose-Einstein condensation.
F. London (1938)
treated the ground state (also known as He~II) as discrete,
and the excited states (also known as He~I) as a continuum,
which is the reason that I refer to it as the binary division approximation.
Even today this binary division
remains the dominant model,
and it is one of the significant differences between my approach
and the conventional approach.

An additional difference is that
the conventional view of Bose-Einstein condensation,
including F. London's (1938) treatment of the $\lambda$-transition,
takes `ground state' to mean ground energy state
and `excited states' to mean excited energy states.
In contrast, I mean ground momentum state and excited momentum states.
For non-interacting bosons the two definitions are equivalent.
Einstein in a letter to Paul Ehrenfest (1924) wrote
\begin{center}
\parbox{7.5cm}{
`From a certain temperature on,
the molecules ``condense'' without attractive forces,
that is, they accumulate at zero velocity.' (Balibar 2014)
}\end{center}
Einstein was referring to ideal bosons.
One can see that the extension of his idea to interacting bosons is ambiguous:
since energy states and momentum states are not equivalent for interacting bosons,
which of the two should be used to describe condensation?
Whereas the conventional interpretation assumes that Einstein meant energy states,
the present paper takes momentum states
to be the natural repository of condensation.

As mentioned,
the present work also disagrees with Einstein's and with F. London's
notion that condensation is solely into the ground state,
whether energy or momentum,
which is another reason that
I do \emph{not} invoke the binary division approximation.
There are detailed reasons, both experimental and mathematical,
for the present interpretation
of Bose-Einstein condensation as being into multiple states,
(Attard  2023 sections~8.2.4, 8.6.6 and 9.2).
To those arguments can now be added the numerical results
of the present QSMD simulations,
which show in the quantum regime
that at any one time multiple momentum states are highly occupied,
that above the condensation transition
the ground momentum state is rarely the most occupied state,
and that below the transition
a minority of the bosons are in the ground momentum state.

This paper differs in several respects
from the previous quantum loop Monte Carlo (QLMC) simulations
of Lennard-Jones $^4$He (Attard 2022, Attard 2023 chapter 8).
That work invoked a permutation loop expansion
and used series of up to seven loops.
It was  valid
on the high temperature side of the $\lambda$-transition,
it used the  binary division approximation,
and it gave only static results.

The present QSMD paper gives dynamic information such as the
momentum time correlation function, the viscosity,
and the individual trajectories of the condensed and the uncondensed bosons.
It uses only pure momentum loops,
which dominate below the $\lambda$-transition.
(Position permutation  loops are important above and at the $\lambda$-transition
(Attard 2022, 2023 chapter 8).)
The present paper dispenses with the binary-division approximation,
and explores momentum state condensation explicitly.

The present paper uses a number of concepts and mathematical analysis
from my previous work.
In some cases it provides numerical confirmation for these earlier predictions.
The present paper  is best read in conjunction with chapters 7, 8, and 9
of the recent edition of my book,
\emph{Entropy Beyond the Second Law}
(Attard 2023 second edition).

%
\section{Analysis and Simulation Algorithm}
\setcounter{equation}{0} \setcounter{subsubsection}{0}
%

\subsection{Analysis for the Static Structure}

Following Attard (2022, 2023 chapters 7 and 8),
consider a system of $N$  identical bosons
interacting with potential energy $U({\bf q})$,
where ${\bf q} = \{{\bf q}_1,{\bf q}_2,\ldots,{\bf q}_N\}$
is the position configuration.
The classical kinetic energy is
${\cal K}({\bf p}) = p^2/2m$,
where ${\bf p} = \{{\bf p}_1,{\bf p}_2,\ldots,{\bf p}_N\}$
is the momentum configuration,
${\bf p}_j$ being the momentum eigenvalue of boson $j$.
The normalized  momentum eigenfunctions for the discrete momentum case
are
$|{\bf p}\rangle = V^{-N/2} e^{-{\bf q}\cdot{\bf p}/\mathrm{i}\hbar}$.
The system has volume $V=L^3$
and the spacing between momentum states is
$\Delta_p = 2\pi \hbar/L$ (Messiah 1961, Merzbacher 1970).
Below the continuum limit will be taken.

An open quantum system is one that exchanges energy
and other conserved variables with its environment.
As a result of entanglement the subsystem becomes decoherent,
and a formally exact expression for
the grand partition function for bosons is
(Attard 2018b, 2021)
\begin{eqnarray}
\Xi^+
& = &
\mbox{TR}' \; e^{-\beta \hat{\cal H}}
\nonumber \\ & = &
\sum_{N=0}^\infty \frac{z^N}{N!}
\sum_{\hat{\mathrm P}} \sum_{\bf p}
\left\langle \hat{\mathrm P} {\bf p}
\left| e^{-\beta \hat{\cal H}} \right| {\bf p} \right\rangle
\nonumber \\ & = &
\sum_{N=0}^\infty \frac{z^N}{N!}
\sum_{\hat{\mathrm P}} \sum_{\bf p}
\int \mathrm{d}{\bf q}\;
\left\langle \hat{\mathrm P} {\bf p} | {\bf q} \right\rangle \,
\left\langle {\bf q}  \left| e^{-\beta \hat{\cal H}}
\right| {\bf p} \right\rangle
\nonumber \\ & \approx &
\sum_{N=0}^\infty \frac{z^N}{N! V^{N}}
\sum_{\hat{\mathrm P}} \sum_{\bf p}
\int \mathrm{d}{\bf q}\; e^{-\beta {\cal H}({\bf q},{\bf p})}
\frac{
\left\langle \hat{\mathrm P} {\bf p} | {\bf q} \right\rangle
}{ \left\langle {\bf p}  | {\bf q} \right\rangle }
\nonumber \\ & = &
\sum_{N=0}^\infty \frac{z^N}{N! V^{N}}
\sum_{\bf p} \int \mathrm{d}{\bf q}\;
e^{-\beta {\cal H}({\bf q},{\bf p})}
\eta^+({\bf q},{\bf p}) .
\end{eqnarray}
The permutation operator is $\hat{\mathrm P}$.
In the penultimate equality the commutation function has been neglected;
the error introduced by this approximation is negligible
in systems that are dominated by long range effects (Attard 2018b,  2021).
The classical Hamiltonian phase space function is
${\cal H}({\bf q},{\bf p}) = {\cal K}({\bf p}) + U({\bf q})$.
The symmetrization function $\eta^+({\bf q},{\bf p})$
is the sum of Fourier factors over all boson permutations.

Note that the quantum state of the open subsystem
is given by the simultaneous specification
of the positions and momenta of its particles,
and that this is a formally exact consequence of the decoherence
of the open quantum subsystem (Attard 2018b, 2021).
The consequences of the lack of simultaneity of position and momentum,
which is a well-known if unrealistic consequence
of the Copenhagen interpretation of quantum mechanics,
are subsumed into the commutation function.
This is neglected in the present paper,
although more generally it can be accounted for (Attard 2018b, 2021, 2023).
This specification of the positions and momenta
is a form of quantum realism that goes further
than the de Broglie-Bohm pilot wave theory,
although it is strictly valid only for open quantum subsystems.

The grand partition function
is just the sum over number of the canonical partition function,
$\Xi^+(z,V,T)  = \sum_{N }  z^N Z^+(N,V,T)$,
with the latter being (Attard 2023 equation~(9.1))
\begin{eqnarray}
\lefteqn{
Z^+(N,V,T)
}  \\ \nonumber
& = &
\frac{1}{N!}\sum_{\hat{\mathrm P}}
 \frac{ 1 }{ V^{N} } \int  \mathrm{d}{\bf q}\;
e^{-\beta U({\bf q})}
\sum_{\bf p}
e^{-\beta p^2 /2m}
e^{{\bf q} \cdot ({\bf p}-{\bf p}') /\mathrm{i}\hbar }
\\ \nonumber
& = &
Z^{+,\mathrm{id}}(N,V,T) \frac{ Q^\mathrm{cl}(N,V,T)}{ V^{N}}  .
\end{eqnarray}
Here the momentum state is ${\bf p} = {\bf n} \Delta_p$,
where ${\bf n} $ is a $3N$-dimensional integer vector,
and the permuted state is ${\bf p}' \equiv \hat{\mathrm P} {\bf p}$.
The classical configuration integral is
$Q^\mathrm{cl}(N,V,T) = \int \mathrm{d}{\bf q}\; e^{-\beta U({\bf q})}$.

The factorization of the momentum terms occurs
when only permutations between bosons in the same momentum state are allowed,
$p_{j\alpha}-p_{j\alpha}' = 0$,
which makes the symmetrization function independent of the position configuration.
This is valid at low temperatures where there is a limited number
of accessible momentum states (Attard 2023 section~9.3).
Under these circumstances the permutation loops
are predominantly momentum loops
in which all particles in a loop are in the same momentum state.
The momentum factor that arises is
\begin{eqnarray}
\lefteqn{
Z^{+,\mathrm{id}}(N,V,T)
} \nonumber \\
& = &
\frac{1}{N!} \sum_{\hat{\mathrm P}}
\prod_{j,\alpha} \sum_{n_{j\alpha}=-\infty}^\infty
e^{-\beta \Delta_p^2 n_{j\alpha}^2/2m}
\delta_{n_{j\alpha},n_{j\alpha}'} .
\end{eqnarray}
This neglects position permutation loops,
which arise from the integration over the momentum states
in which bosons in the same state form a set of measure zero,
and which do not necessarily contribute zero upon integration
because of the interaction potential in the integrand
(Attard 2023 section~8.5).
However, when pure momentum loops give the dominant contribution,
as is assumed here,
then all the wave function symmetrization effects are contained
within this momentum factor.
Because of the factorization, the momentum integral
is the same as that for non-interacting bosons,
and it is directly connected with
the known grand canonical ideal bosons results
(F. London 1938, Pathria 1972 chapter~7, Attard 2023 section~8.2).
The present analysis differs from that analysis in three ways:
First it avoids the binary division continuum approximation,
with the focus being on the actual occupancy of the discrete momentum states.
Second, the relationship between fugacity and number,
$\overline z(N,V,T)$,
is different for interacting and for ideal bosons.
And third, the boson dynamics depend upon their interactions.

It must be emphasized that this factorization is predicated
upon the neglect of the commutation function,
and upon the explicit restriction to pure momentum permutation loops.
The latter can be expected to be valid below the $\lambda$-transition,
but the neglect of position loops limits its utility at higher temperatures
(Attard 2023 section~8.5).

The ideal boson partition  function above
contains the Kronecker-$\delta$, $\delta_{n_{j\alpha},n_{j\alpha}'}$.
In the momentum configuration ${\bf p} = {\bf n}\Delta_p$,
there are
\begin{equation}
N_{\bf a}({\bf p}) = \sum_{j=1}^N \delta_{{\bf p}_j, {\bf a}}
\end{equation}
bosons in the single particle momentum state ${\bf a}$.
Hence there are (Attard 2021 equation~(2.20))
\begin{equation}
\chi_{\bf n}^+
=
\sum_{\hat{\mathrm P}}
 \langle\hat{\mathrm P} {\bf n} | {\bf n} \rangle
=
 \prod_{\bf a} N_{\bf a}({\bf n})!
\end{equation}
non-zero permutations of the bosons in the system.
Therefore the permutation entropy associated with
a momentum configuration is
\begin{equation}
S^\mathrm{perm}({\bf p})
= k_\mathrm{B} \ln \chi_{\bf n}^+
= k_\mathrm{B}  \sum_{\bf a} \ln N_{\bf a}({\bf p})! .
\end{equation}
(This is the internal entropy of the point in phase space,
which is in addition to the external reservoir entropy
that is discussed shortly.)
With this the ideal boson partition function becomes
\begin{equation}
Z^{+,\mathrm{id}}(N,V,T)
=
\frac{1}{N!} \sum_{\bf p}
e^{-\beta {\cal K}({\bf p})}
e^{S^\mathrm{perm}({\bf p})/ k_\mathrm{B}} ,
\end{equation}
where the classical kinetic energy function is
${\cal K}({\bf p}) = \sum_{j=1}^N p_j^2/2m = p^2/2m$.
Taking the continuum momentum limit (discussed below),
the full partition function is
\begin{eqnarray}
\lefteqn{
Z^{+}(N,V,T)
} \nonumber \\
& = &
\frac{1}{N! \Delta_p^{3N} V^N }
\int \mathrm{d}{\bf p}  \int \mathrm{d}{\bf q} \;
e^{S^\mathrm{perm}({\bf p})/ k_\mathrm{B}}
e^{-\beta {\cal K}({\bf p})}
e^{-\beta U({\bf q})}
\nonumber \\ & = &
\frac{1}{N! h^{3N}} \int \mathrm{d}{\bf \Gamma} \;
e^{S^\mathrm{perm}({\bf p})/ k_\mathrm{B}}
e^{S^\mathrm{r}({\bf \Gamma})/ k_\mathrm{B} } .
\end{eqnarray}
Here the reservoir entropy is the usual Maxwell-Boltzmann factor
\begin{equation}
S^\mathrm{r}({\bf \Gamma}) = \frac{-{\cal H}({\bf \Gamma})}{T} ,
\end{equation}
where the Hamiltonian function of classical phase space is
the sum of the kinetic and potential energies,
${\cal H}({\bf \Gamma}) = {\cal K}({\bf p}) + U({\bf q})$.

One can identify from this
that the canonical equilibrium quantum probability density
of a point in classical phase space is proportional to
the exponential of the total entropy,
\begin{equation}
\wp({\bf \Gamma}) =
\frac{1}{Z'} e^{S({\bf \Gamma})/ k_\mathrm{B} },
\;\;
S({\bf \Gamma}) = S^\mathrm{perm}({\bf p}) + S^\mathrm{r}({\bf \Gamma}) .
\end{equation}
The relation between entropy and probability
is to be expected on very general grounds
(Attard 2002, Attard 2012, Attard 2023 chapter~1).

This expression for the total entropy
is the same as in the treatment on the far-side of the $\lambda$-transition
in Attard (2023 section~9.3).
It includes only permutations of bosons in the same momentum state
(pure momentum loops).
It precludes the position loops that
dominate on the high-temperature side of the $\lambda$-transition,
where their increase in number and size gives the marked increase
in the heat capacity leading up to the $\lambda$-transition
(Attard 2023 section~8.5).

Below I shall discuss the relation between the continuum limit
and the occupancy of the quantized momentum states
with regard to the dynamics of the system.
For the continuum momentum configuration
${\bf p} =\{{\bf p}_1,{\bf p}_2, \ldots ,{\bf p}_N\}$,
the simplest definition  of the discrete occupancy
of the single-particle momentum state ${\bf a} = {\bf n} \Delta_p$
is
\begin{eqnarray} \label{Eq:DisctsOcc}
N_{\bf a}({\bf p})
& = &
\sum_{j=1}^N \prod_{\alpha=x,y,z}
\Theta\big( p_{j\alpha} - ( a_{\alpha} - \Delta_p/2) \big) \,
\nonumber \\ &  & \mbox{ } \times
\Theta\big( ( a_{\alpha} + \Delta_p/2) - p_{j\alpha} \big) ,
\end{eqnarray}
where the Heaviside unit step function appears.

\subsection{Decoherent Quantum Equations of Motion}

As mentioned in the introduction,
the focus of this paper is on the dynamics of superfluidity.
The above analysis gives the state of an open quantum system of bosons
as a point in classical phase space.
In the present subsection
molecular-level equations of motion
for the decoherent quantum system are derived.

The decoherent quantum velocity in phase space is (Attard 2023 section~9.4.4)
\begin{equation} \label{Eq:dotG0}
\dot{\bf \Gamma}^0 = -T \nabla^\dag S({\bf \Gamma}) .
\end{equation}
This is a major new result for the time evolution of open quantum systems.
A point in classical phase space is
${\bf \Gamma} \equiv \{ {\bf q}, {\bf p}\}$,
the gradient operator is $\nabla \equiv\{ \nabla_q, \nabla_p \}$,
and its conjugate is $\nabla^\dag \equiv\{ \nabla_p, {- \nabla_q} \}$.
The superscript 0 signifies the non-thermostatted decoherent trajectory.

The rationale for this expression is that it ensures
that time averages equal phase space averages
(Attard 2002 section~5.1.3,
2012 section~7.2.3,
2023 section~5.2.3).
Hamilton's  equations of motion,
which are the classical adiabatic equations of motion,
is this with $S \Rightarrow S^\mathrm{r} \equiv - {\cal H}/T$
(ie.\ neglecting the permutation entropy of the quantized momentum states).
Since the permutation entropy is an even function of the momenta,
$S^\mathrm{perm}(-{\bf p}) = S^\mathrm{perm}({\bf p})$,
as is the reservoir entropy,
these  decoherent quantum equations of motion are time reversible
(cf.\ Attard 2023 section~5.4.2).
These equations are incompressible, $\nabla \cdot \dot{\bf \Gamma}^0  = 0$,
so that volume elements are conserved on a decoherent quantum  trajectory
(Attard 2002 section~5.1.5,
2012 section~7.2.4,
2023 section~5.1.2).
The decoherent rate of change of entropy vanishes,
$ \dot{\bf \Gamma}^0 \cdot \nabla S  = 0$,
as can be seen by inspection.
Putting these together
these equations of motion ensure that
the quantum probability density in phase space,
$\wp({\bf \Gamma}) = Z'^{-1}  e^{S({\bf \Gamma})/k_\mathrm{B}}$,
is a constant of the decoherent quantum  motion.

In the general case, $S = S^\mathrm{r} + k_\mathrm{B}\ln(\eta \omega)$.
In this paper the commutation function is neglected,
$\omega({\bf \Gamma}) = 1$.
In general the symmetrization function gives the permutation entropy,
$S^\mathrm{perm}({\bf p}) = k_\mathrm{B}\ln \eta({\bf \Gamma})$.
In this paper this is approximated by pure momentum loops.

In this paper the only position dependence for the entropy
comes from the potential energy contribution to the reservoir entropy.
Hence for a homogeneous system in which the potential energy
is translationally invariant,
the total momentum is a constant of the motion
on this decoherent quantum trajectory.

Of course, it is a formally exact statistical requirement
that the equilibrium probability density be stationary
on the decoherent quantum trajectory.
Note that the decoherent quantum equations of motion
do \emph{not} maximize the entropy.
Rather, they maintain it.
The dissipative term (next) in the equations of motion will tend
to increase the entropy in the event that the current point in phase space
is an unlikely one (appendix~\ref{Sec:Stab}).
The superscript 0 is used to distinguish the decoherent quantum term
from the stochastic dissipative thermostat contribution.
The decoherent quantum equations of motion,
as well as the stochastic dissipative terms,
are for a statistical quantum system that is open to the environment.

The decoherent quantum evolution to linear order in the time step is
\begin{equation}
{\bf \Gamma}^0(t+\tau|{\bf \Gamma}(t),t)
=
{\bf \Gamma}(t)+ \tau \dot{\bf \Gamma}^0(t).
\end{equation}
The second entropy formulation of non-equilibrium thermodynamics
shows that the general form for
the equations of motion is necessarily stochastic and dissipative
(Attard 2012 sections~3.6 and 7.4.5, 2023 section~5.4).
Hence the generic equations of motion in classical phase space are
(Attard 2023 section~9.4.4)
\begin{equation} \label{Eq:SDdQ-EoM}
{\bf \Gamma}(t+\tau)
=
{\bf \Gamma}^0(t+\tau|{\bf \Gamma}(t),t) + {\bf R}_p(t).
\end{equation}
The thermostat contribution has only momentum components,
${\bf R}_p(t) = \overline {\bf R}_p(t) + \tilde{\bf R}_p(t)$.
This has been proven in the classical case (Attard 2012 sections~3.6 and 7.4.5)
and it is assumed that it carries over to the present quantum case.
The dissipative force is
\begin{equation}
\overline {\bf R}_p(t) =
\frac{\sigma^2}{2k_\mathrm{B}}   \nabla_p S({\bf \Gamma}(t))  ,
\end{equation}
and the stochastic force satisfies
\begin{equation}
\langle \tilde {\bf R}_p(t) \rangle = {\bf 0} ,
\mbox{ and }
\langle \tilde {\bf R}_p(t) \tilde {\bf R}_p(t') \rangle
= \sigma^2\delta_{t,t'}\mathrm{I}_{pp} .
\end{equation}
As these terms are derived from a linear expansion
of the transition probability over a time step
(Attard 2012 sections~3.6 and 7.4.5, 2023 section~5.4),
the thermostat should represent a small perturbation
on the decoherent quantum trajectory.
The results should be independent of the stochastic parameter,
provided that it is large enough to compensate the inevitable numerical error
that arises from the first order equation of motions  over the finite time step.
Further, as is discussed in the results section~\ref{Sec:Cv},
as $\sigma$ is made smaller the classical part of the trajectory
tends toward that of an adiabatic (microcanonical) system,
and the energy fluctuations and the heat capacity
are reduced from their canonical values.

In appendix~\ref{Sec:Stab} it is proven that the equilibrium probability density
is stable during its evolution according to these equations of motion.

In detail the equations of motion are to first order in the time step
\begin{eqnarray}
q_{j\alpha}(t+\tau)
& = &
q_{j\alpha}(t) - \tau T \partial_{p,j\alpha} S(t)
\nonumber \\
p_{j\alpha}(t+\tau)
& = &
p_{j\alpha}(t) + \tau T \partial_{q,j\alpha} S(t)
\nonumber \\ && \mbox{ }
+ \frac{\sigma^2}{2k_\mathrm{B}} \partial_{p,j\alpha} S(t)
+ \tilde R_{p,j\alpha}(t).
\end{eqnarray}
Second order equations of motion are derived in appendix~\ref{Sec:2ndOEoM}.
Here and below, $j=1,2,\ldots,N$ and $\alpha =x,y,z$.
Of course $\partial_{p,j\alpha} S^\mathrm{r}(t) = - p_{j\alpha}/mT$,
and $\partial_{q,j\alpha} S(t)
= \partial_{q,j\alpha} S^\mathrm{r}(t)
= - \partial_{q,j\alpha} U(t)/T
= f^0_{j\alpha}(t)/T$,
which are the classical velocity
and the classical adiabatic force (divided by the temperature),
respectively.
Notice how the permutation entropy contributes to the
decoherent quantum evolution of the positions.
This breaks the classical nexus between momentum
and the rate of change of position,
and it gives a non-local contribution to the latter.

These equations of motion with continuous momentum
give the transition rate for the quantized momentum states.
They ensure the exact quantum equilibrium probability density
in classical phase space with quantized momentum states,
which
itself is the exact formulation of a quantum statistical system
that allows position and momentum to be simultaneously specified.
(Actually,
the formula $\dot{\bf \Gamma}^0 = - T \nabla^\dag S({\bf \Gamma})$ is exact;
the present implementation invokes two approximations for the entropy,
namely the neglect of the commutation function,
and the restriction to pure momentum permutation loops.)
The continuous momentum can be seen as the realization of the bosons
between momentum measurements,
and there is no reason
that the values should not interpolate momentum eigenvalues.
The decoherent quantum equations of motion could be interpreted as
the realistic manifestation of Schr\"odinger's equation
for an open quantum system.

For a macroscopic system,
the spacing between quantum levels is immeasurably small,
and so for all phase space functions of practical interest
the continuum limit may be taken,
the only exceptions being
the momentum state occupancy and the permutation entropy.
It remains to obtain an expression for the gradient of the permutation entropy,
$\partial_{p,j\alpha} S^\mathrm{perm}({\bf p}(t)) $.

\subsection{Continuous Occupancy}

In the formulation of the preceding subsection,
the momentum ${\bf p} = \{ {\bf p}_1,{\bf p}_2,\ldots,{\bf p}_N\}$
is a continuous variable that is not quantized.
Although it is trivial to find from this
the occupation of discrete states and the permutation entropy,
equation~(\ref{Eq:DisctsOcc}),
the equations of motion require
the momentum gradient of the permutation entropy,
and this is a  $\delta$-function at the boundaries
of the occupied states (appendix~\ref{Sec:DiscOccTraj}).
It is not  numerically feasible to compute this directly.

This conundrum is resolved in the present paper
by defining a continuous occupancy and permutation entropy,
combined with umbrella sampling.
Two different formulations for the fractional occupancy are used.
If boson $j$ is actually in the state ${\bf a}_j$,
$p_{j\alpha} \in (a_{j\alpha} - \Delta_p/2 , a_{j\alpha} + \Delta_p/2)$,
then the first formulation
defines the fraction of it associated with the state ${\bf a}_j$
in the direction $\alpha$ to be
\begin{subequations}
\begin{equation} \label{Eq:poly}
\phi_{j\alpha} \equiv 1 - [2(p_{j\alpha} - a_{j\alpha})/\Delta_p]^{\kappa}/2 ,
\;\; \kappa=2,4,\ldots
\end{equation}
One sees that  $\phi_{j\alpha} \in [1/2,1]$.
The even positive integer $\kappa$ is chosen with a fixed value;
the larger this is,
the greater is the share of the occupation weight given to the state ${\bf a}_j$,
and the closer to the boundary of the state does the boson have to be
in order to contribute significantly to the occupancy
of the nearest neighboring states.

The second formulation defines the fraction to be
\begin{equation}  \label{Eq:hyp}
\phi_{j\alpha} \equiv
\frac{1}{2}
- \frac{s_{j\alpha}}{2}
\tanh[2\kappa \{p_{j\alpha}- (a_{j\alpha}+a_{j\alpha}')/2 \}/\Delta_p] .
\end{equation}
\end{subequations}
Here $s_{j\alpha} \equiv \mbox{sign}(p_{j\alpha} - a_{j\alpha})$
tells which half of the momentum state the boson is in,
and $a_{j\alpha}' = a_{j\alpha} + \Delta_p s_{j\alpha}$
is the closest neighbor state to boson $j$ along the $p_\alpha$ axis.
Hence the argument of the hyperbolic tangent
tells the distance to the closest boundary.
Again the fraction is equal to one half at the boundary,
and  toward the center of the state
it approaches $0.5 + 0.5 \tanh \kappa \approx 1$,
depending on how large $\kappa$ is.

With these the continuous occupancy
of the single particle momentum state ${\bf a}$
for the continuous momentum configuration ${\bf p}$ is defined as
\begin{eqnarray}
\lefteqn{
{\cal N}_{\bf a}({\bf p})
}  \\
& = &
\sum_{j=1}^N
\Big\{
\phi_{jx}\phi_{jy}\phi_{jz}\delta_{{\bf a},{\bf a}_j}
+
\overline \phi_{jx}  \overline\phi_{jy} \overline\phi_{jz}
\delta_{{\bf a},{\bf a}_j'}
\nonumber \\ && 
+
\overline \phi_{jx} \phi_{jy}\phi_{jz}
\delta_{a_x, a_{jx}'} \delta_{a_y, a_{jy}} \delta_{a_z, a_{jz}}
+ 2 \mbox{ similar terms}
\nonumber \\ && 
+
\overline \phi_{jx} \overline \phi_{jy} \phi_{jz}
\delta_{a_x, a_{jx}'} \delta_{a_y, a_{jy}'} \delta_{a_z, a_{jz}}
+ 2 \mbox{ similar terms}
\Big\},\nonumber
\end{eqnarray}
where $\overline \phi = 1-\phi$.
This formulation distributes
the unit weight of each boson
over its own and seven neighboring states.
The closer the boson is to the center of its momentum state,
the more weight it contributes to its own state.
As mentioned, the larger $\kappa$ is,
the thinner is the boundary region for significant weight sharing,
and  the more the gradient  approaches a $\delta$-function.
If the boson is at the boundary between two momentum states,
exactly half of the respective one-dimensional part of the weight
is contributed to each.

With  this continuous (ie.\ real number) occupancy,
the permutation entropy of a momentum configuration is
\begin{equation}
S^\mathrm{perm}({\bf p})
=
k_\mathrm{B} \sum_{\bf a} \ln \Gamma({\cal N}_{\bf a} + 1 ) ,
\end{equation}
where $\Gamma(z+1) = z!$ is the Gamma function
(Abramowitz and Stegun 1972 chapter~6).

Now $\phi_{j\alpha}$ is the $\alpha$ component of the weight of boson $j$
in the state ${\bf a}_j$,
and $1-\phi_{j\alpha}$ is that in the neighboring state
${\bf a}_j' = {\bf a}_j+ {\bf s}_{j}\Delta_p $.
Changing ${\bf p}_{j}$ changes ${\bm \phi}_j$
and hence ${\cal N}_{{\bf a}_j}$ and the seven ${\cal N}_{{\bf a}'_j}$.
With this the components of the gradient of the permutation entropy are
\begin{eqnarray}
\lefteqn{
\nabla_{p,j\alpha} S^\mathrm{perm}({\bf p})
} \nonumber \\
& = &
k_\mathrm{B}
\sum_{ r_{jx}=0}^1
\sum_{ r_{jy}=0}^1
\sum_{ r_{jz}=0}^1
\frac{\partial \ln \Gamma( {\cal N}_{{\bf a}_j''} + 1 )
}{ \partial {\cal N}_{{\bf a}_j''} }
\;
\frac{\partial \Phi_{j}^{({\bf r}_{j})}  }{\partial p_{j\alpha}} ,
\nonumber \\ && \mbox{ }
{\bf a}_j'' \equiv {\bf a}_j
+ \sum_{\gamma=x,y,z} r_{j\gamma} s_{j\gamma} \Delta_p \widehat{\bm \gamma} .
\end{eqnarray}
Here has been defined
\begin{eqnarray}
\Phi_{j}^{({\bf r}_{j})}
& \equiv &
\Phi_{jx}^{(r_{jx})} \Phi_{jy}^{(r_{jy})} \Phi_{jz}^{(r_{jz})},
 \\ && \mbox{ }\nonumber
\Phi_{j\alpha}^{(r_{j\alpha})}
\equiv
\overline r_{j\alpha} \phi_{j\alpha} + r_{j\alpha} \overline\phi_{j\alpha}  ,
\end{eqnarray}
where $\overline r_{j\alpha} = 1 - r_{j\alpha}$.
This is the contribution from boson $j$
to the occupancy of the state ${\bf a}_j''$,
and one has
$\partial_{p,j\alpha} {\cal N}_{{\bf a}_j''}
= \partial_{p,j\alpha} \Phi_{j}^{({\bf r}_{j})}$.
The $x$-component of the required derivative is
\begin{eqnarray}
\frac{\partial  \Phi_{j}^{({\bf r}_{j})} }{\partial p_{jx}}
& = &
(1-2r_{jx})\, (\partial_{p_{jx}} \phi_{jx})
 \Phi_{jy}^{(r_{jy})}\Phi_{jz}^{(r_{jz})} ,
\end{eqnarray}
and similarly for the other two components.

The continuous occupancy formulation is equivalent
to sampling phase space with an umbrella weight.
One should correct for this by taking the average over the trajectory to be
\begin{equation}
\langle f({\bf \Gamma}) \rangle_{N,V,T}
=
\frac{
\sum_n f({\bf \Gamma}(t_n))
\prod_{\bf a} \frac{N_{\bf a}(t_n)!}{\Gamma({\cal N}_{\bf a}(t_n)+1)}
}{
\sum_n
\prod_{\bf a} \frac{N_{\bf a}(t_n)!}{\Gamma({\cal N}_{\bf a}(t_n)+1)}
}.
\end{equation}
Umbrella sampling is of course a formally exact statistical technique
(ie. confidence that the average value is exact increases
with increasing number of samples).
In practice however,
the computational efficiency
(ie.\ the statistical error for a fixed number of time steps)
can be quite sensitive to the choice of umbrella function and parameters.
In the long run the average value is not affected by such a choice,
but the confidence in the value is.

The ratio of the actual weight to the umbrella weight
can be quite large,
(the present fractional occupancy functions systematically underestimate
the occupancy of highly occupied states)
and it increases with system size.
The larger the ratio
the lower the statistical accuracy for a given length of simulation.
In both formulations of the fractional occupancy,
the umbrella weight ratio approaches unity with increasing $\kappa$.
However, there are practical computational limits to how large
$\kappa$ can be,
including the cost  of evaluating the fractional occupancy function,
and its accuracy.
Also large permutation entropy gradients,
which scale with $\kappa$ and the proximity to a momentum state boundary,
necessitate a smaller time step than
would suffice for the classical equations of motion.

It was found in practice that the fractional occupancy underestimated
the actual occupancy of highly occupied states,
and overestimated that of few occupied states.
The former is the more significant effect,
and it leads to the  umbrella weight ratio
being significantly greater than unity.
To ameliorate this, most runs reported below were performed
with the continuous occupancy defined as
$\tilde {\cal N}_{\bf a} = c {\cal N}_{\bf a}$,
with $c =$ 1.02--1.04,
depending on the temperature and the function used for the fractional occupancy.
(The value was determined by making
the average of the ratio of the actual weight to the umbrella weight
close to unity in some smaller equilibration runs.)
The first derivative of the permutation entropy
given above should be multiplied by
$\partial_{{\cal N}_{\bf a}}  {\tilde{\cal N}_{\bf a}} = c$,
since
$\nabla_p S^\mathrm{perm}
= k_\mathrm{B} \sum_{\bf a}
[\partial_{\tilde{\cal N}_{\bf a}} \ln \tilde{\cal N}_{\bf a}!]
[\partial_{{\cal N}_{\bf a}}  {\tilde{\cal N}_{\bf a}}]
\nabla_p {\cal N}_{\bf a}$.
The second derivative given in appendix~\ref{Sec:2ndOEoM}
similarly contains terms linear and quadratic in $c$.

\subsection{Computational Details}

The quantum stochastic molecular dynamics (QSMD) algorithm
has much in common with earlier classical SMD versions
(Attard 2012 section~11.1).
Periodic boundary conditions and the minimum image convention
were used in position space.
Both first and second order equations of motion were used,
with the latter taking about three times longer to evaluate at each time step
but enabling a ten times larger time step.
The time step and thermostat parameters chosen were sufficient
in the classical case to yield the known exact classical kinetic energy
and fluctuations therein to within a few per cent.
The Lennard-Jones pair potential was used
$u_\mathrm{LJ}(q_{jk}) = 4\epsilon_\mathrm{LJ}[(\sigma_\mathrm{LJ}/q_{jk})^{12}
-(\sigma_\mathrm{LJ}/q_{jk})^{6}]$,
with potential cut-off of $R_\mathrm{cut} = 3.5 \sigma_\mathrm{LJ}$.
The Lennard-Jones parameters for helium that were used were
$ \epsilon_\mathrm{He} = 10.22 k_\mathrm{B}$\,J
and $\sigma_\mathrm{He} = 0.2556$\,nm
(van Sciver 2012).
A spatially based small cell neighbor table was used.
The number density was $\rho = N/L^3$,
and the spacing between momentum states was
$\Delta_p = 2\pi \hbar/L$.
For most of the simulations $N=1,000$.
Some tests were performed with up to $N=10,000$ in the classical case;
in the quantum case the performance of the umbrella sampling
deteriorated with increasing system size.
The simulations were for a homogeneous system
at the liquid saturation density at each temperature (Attard 2023 table~8.3).

The simulation was broken into 10 blocks,
and the fluctuation in the block averages was used to estimate
the statistical error.
Repeat runs were also used to estimate independently the statistical error.
Generally a single run of the program
for $N=1000$ consisted of $5 \times10^5$ time steps.
The internal statistical error for the occupancies
estimated from the fluctuations
in the averages over each of 10 blocks of this
was quite often much smaller than the statistical error
estimated from the fluctuations in the averages in a series of consecutive runs.
This says that the occupancies over consecutive blocks
were substantially correlated,
which gives an indication of how large and long-lived
are the fluctuations in momentum state occupancies.
Since the start point of each run was the end point of the previous run,
there is no guarantee that correlations
do not cause a non-negligible underestimate
in the occupancy error obtained from consecutive runs.
The statistical error reported below
is generally the larger of the various estimates.

The author's experience with classical systems and with Monte Carlo simulations
has led him to conclude that in the quantum regime
the present system displays unusually large
and long-lived fluctuations, and it is very slow to equilibrate.

The Gamma function was evaluated
using small and large argument expansions
(Abramowitz and Stegun 1972 equations~(6.1.36) and (6.1.41)).
These are readily differentiated.

The viscosity was obtained from
the momentum-moment--momentum-moment  time-correlation function
(Attard 2012 equation~(9.117)),
\begin{equation} \label{Eq:eta(t)}
\eta_{xz}(t)
=
\frac{1}{2 V  k_\mathrm{B} T}
\int_{-t}^{t} \mathrm{d}t'\,
\left< \dot P_{xz}^0({\bf \Gamma})
\dot P_{xz}^0({\bf \Gamma}(t'|{\bf \Gamma},0))
\right> .
\end{equation}
This is often called a Green-Kubo expression (Green 1954, Kubo 1966).
In fact it was Onsager who first gave
the relationship between the transport coefficients
and the time correlation functions (Onsager 1931),
and the present author's own derivation owes much to this (Attard 2012).
The first $z$-moment of the $x$-component of momentum is
$P_{zx}({\bf \Gamma}) = \sum_{j=1}^N z_j p_{xj}$.
In the classical case $\dot P_{zx}^0 = \dot P_{xz}^0$ (see next).
The author's classical derivation of the Green-Kubo relations
via the second entropy assumes a Markov regression regime
in which plateau region $\eta(t)$ is independent of $t$.
In practice $\eta(t)$ varies most slowly with $t$ at its maximum,
which is taken as `the' value of the viscosity.
This is justified by the fact that
the curvature of the plateau region decreases with increasing system size,
and it can be expected to become flat in the thermodynamic limit.
Obviously the maximum value can be sensitive to statistical errors,
particularly since these grow with $t$.
The results below for $\eta(t)$ allow the reader
to judge for themselves the best estimate of the viscosity.

The trajectory ${\bf \Gamma}(t'|{\bf \Gamma},0)$
is meant to be the decoherent quantum one,
but the results below use the thermostatted trajectory.
The results do not appear sensitive
to the value of the thermostat parameter $\sigma$.
Perhaps more significant is that the trajectory
is the one generated
using the continuous occupancy for the permutation entropy,
and no correction for this is made beyond
applying the ratio of the actual weight to the umbrella weight
at the start of the trajectory during the phase space averaging.
One should really apply a trajectory weight correction that is the product
of the umbrella to actual transition probability weights
for each point on the trajectory.

The above is one of several expressions for the viscosity
(Attard 2012 equations~(9.116) and (9.117))
that are nominally equivalent,
but which differ in their statistical behavior.
Of these, the present one that depends only upon  the rate of change
of the momentum moment is unique in being suitable for systems
with periodic boundary conditions
and the minimum image convention.
(The reason that the other expressions don't work is
that they depend upon the momentum moment rather than its rate of change,
and when bosons enter and leave the central simulation cell
over the course of the trajectory the momentum moment is messed up.)
The decoherent quantum rate of change of the first momentum moment is
\begin{eqnarray} \label{Eq:.P0zx}
\dot P_{xz}^0({\bf \Gamma})
& = &
\dot{\bf \Gamma}^0 \cdot \nabla P_{xz}({\bf \Gamma})
\nonumber \\ & = &
\sum_{i=1}^N
\left[ \frac{ {p}_{zi} p_{xi} }{m}
- T p_{xi} \nabla_{p,zi} S^\mathrm{perm}({\bf p})
\right]
\nonumber \\ &  & \mbox{ }
- \sum_{i<j}^N u'(q_{ij}) \frac{[z_i-z_j] [x_{i}-x_{j}]}{q_{ij}} .
\end{eqnarray}
This is suitable when the potential energy
consists solely of central pair terms.
The first term is
the contribution to the rate of momentum moment change
due to molecular diffusion
(the first term of which is classical),
and the second term is the contribution from intermolecular forces.
This expression depends upon the relative separations
rather than the absolute positions,
which is why the value of the viscosity
based upon it is independent of the system volume
when periodic boundary conditions
and the minimum image convention are applied.

Compared to the classical expression
(Attard 2012 equation~(9.119)),
this contains the additional decoherent quantum term,
$\dot q^{0,\mathrm{qu}}_{iz}  p_{xi}
= - T p_{xi} \nabla_{p,iz} S^\mathrm{perm}({\bf p})$,
which is directly responsible for the reduction in the viscosity
of the quantum liquid in the condensed regime
(section~\ref{Sec:Visco}).
Unlike in the classical case,
this quantum contribution is \emph{not} symmetric,
so that
$\dot P_{zx}^0 \ne \dot P_{xz}^0$.
(This means that there are six independent estimates of the viscosity
in each simulation.)
The viscosity tensor, which is this multiplied by its evolved value,
integrated, and averaged,
is expected to be symmetric for a homogeneous system.

The results below are presented in dimensionless form:
the temperature is $T^* = k_\mathrm{B}T/\epsilon_\mathrm{LJ}$,
the number density is $\rho^* = \rho \sigma_\mathrm{LJ}^3$,
the time step is $\tau^* = \tau /t_\mathrm{LJ}$,
the unit of time is
$t_\mathrm{LJ} = \sqrt{m\sigma_\mathrm{LJ}^2/\epsilon_\mathrm{LJ}}$,
the variance of the stochastic force is
$\sigma^* =  \sigma /\sqrt{mk_\mathrm{B}T \tau/t_\mathrm{LJ}}$,
the momentum  is $p_x^* = p_x /\sqrt{mk_\mathrm{B}T}$,
and the viscosity is
$\eta^* = \eta \sigma_\mathrm{LJ}^3/\epsilon_\mathrm{LJ} t_\mathrm{LJ}$.

%
\section{Results}
\setcounter{equation}{0} \setcounter{subsubsection}{0}


\begin{table*}[tb]
\caption{ \label{Tab:t1}
Stochastic molecular dynamics
results for Lennard-Jones $^4$He at liquid saturation density
($N=1000$, $\tau^* = 10^{-4}$, second order equations of motion).
Msteps is millions of time steps,
$| \tilde \Delta {\cal K}/\Delta^0{\cal K} |$
is the ratio of the stochastic to the decoherent quantum change
in kinetic energy per time step,
{\tt avocc} is $N$ divided by the average number of occupied states,
and {\tt maxocc} is the average number in the maximally occupied state.
For the quantum cases,
the hyperbolic fractional occupancy, Eq.~(\ref{Eq:hyp}),
was used with $\kappa=11$ and $c =$ 1.02--1.04.
The number in parentheses is twice the standard error
of the final digit or digits, which gives the 95\% confidence interval.
}
\begin{center}
\begin{tabular}{c c c c c c c c c c c c c c c }
\hline\noalign{\smallskip}
$T^*$ & $\rho^*$ & $\rho \Lambda^3$ & Msteps  & $\sigma^*$ &
$\beta {\cal K}/N$ & $\beta U/N$ &
$| \tilde \Delta_{\cal K}/\Delta^0_{\cal K} |$  &
{\tt avocc} & {\tt maxocc} & $N_{\bf 0}$ &
$C_V/Nk_\mathrm{B}$ & $\eta^*(0.5)$ & $\eta^*(1.5)$
\\
\hline 
\multicolumn{13}{c}{quantum} \\ 
0.70 & 0.847 & 1.76 & 2.5 & 1.0 &
0.93(1) & -8.77(2) & 2.4 & 1.98(2) & 33(5) & 26(4) & 1.3(4) &
2.8(21) & 3.9(37) \\
0.65 & 0.868 & 2.02 & 4.5 & 1.0 &
0.86(2) & -9.75(2) & 3.0 & 2.18(3) & 52(10) & 45(12) & 1.2(5) &
0.93(91) & 2.6(13) \\ 

0.60$^\ddag$  & 0.887 & 2.33 & 7 & 0.2 &
0.84(1) & -10.832(7) & 0.13 & 2.35(3) & 62(7) & 53(8) & 1.8(5) &
2.7(7) & 3.9(18) \\ 
0.60 & 0.887 & 2.33 & 2.5 & 1.0 &
0.77(1) & -10.86(1) & 2.6 & 2.48(5) & 86(11) & 86(12) & 1.3(5) &
1.3(12) & 2.1(19) \\ 

0.55$^\ddag$ & 0.905 & 2.70 & 5 & 0.2 &
0.75(3) & -12.12(2) & 0.09 & 2.67(5) & 96(19) & 94(21) & 1.3(5) &
1.7(7) & 0.7(25) \\ 
0.55 & 0.905 & 2.70 & 3.5 & 0.2 &
0.77(2) & -12.11(2) & 0.13 & 2.61(7) & 83(23) & 80(25) & 1.4(9) &
2.6(13) & 4.7(33) \\ 
0.55 & 0.905 & 2.70 & 5 & 1.0 &
0.68(1) & -12.15(2) & 3.1 & 2.81(5) & 147(23) & 147(23) & 2.1(11) &
3.9(12) & 7.8(28) \\ 
\hline
\multicolumn{13}{c}{classical} \\ 
0.70 & 0.847 & 1.76 & 2 & 1.0 &
1.501(6) & -8.748(6) & 2.1 & 1.302(2) & 5.39(2) & 1.75(3) & 2.5(2) &
3.05(32) & 3.73(51) \\ 
0.65 & 0.868 & 2.02 & 1.5 & 0.2 &
1.51(2) & -9.71(2) & 0.08 & 1.34(1) & 5.70(7) & 1.97(4) & 1.9(4) &
3.11(35) & 3.8(6) \\ 
0.65 & 0.868 & 2.02 & 1 & 0.5 &
1.50(3) & -9.72(2) & 0.50 & 1.35(1) & 5.73(7) & 1.99(4) & 2.4(9) &
3.03(44) & 3.7(12) \\ 
0.60$^\dag$ & 0.887 & 2.33 & .4--.6 & 0.2 &
1.42(2) & -10.89(1) & 0.17 & 1.43(1) & 7.47(7) & 2.51(7) & 0.03(6) &
3.81(99) & 5.0(23) \\ 
0.60 & 0.887 & 2.33 & 3 & 0.2 &
1.53(2) & -10.79(2) & 0.08 & 1.383(1) & 6.00(7) & 2.21(6) & 0.08(1) &
4.02(51) & 5.1(10)  \\ 
0.60 & 0.887 & 2.33 & 1.5 & 0.5 &
1.51(1) & -10.81(2) & 0.5 & 1.392(5) & 6.06(4) & 2.30(4) & 2.3(2) &
4.16(58) & 6.1(14) \\ 
0.55 & 0.905 & 2.70 & 2 & 1.0 &
1.501(8) & -12.12(1) & 2.0 & 1.456(4) & 6.53(3) & 2.66(4) & 2.8(6) &
4.56(40) &  6.9(12) \\ 
\hline
\end{tabular} \\
$^\dag$$N=5,000$,\;
$^\ddag$$\tau^* = 0.5 \times 10^{-4}$ 
\end{center}
\end{table*}


Table~\ref{Tab:t1}
shows quantum stochastic  molecular dynamics (QSMD) results
at four points on the Lennard-Jones $^4$He liquid saturation curve.
Quantum loop Monte Carlo (QLMC) simulations,
give the $\lambda$-transition in Lennard-Jones $^4$He  at about
$T^* = 0.625$ (Attard 2023 chapter~8).
This is driven by the divergence
in the position permutation loops (Attard 2023 section~8.4, 8.6, 9.2),
which are neglected here.
The present QSMD simulations invoke pure momentum permutation loops
via momentum state occupancy,
which results from wave function symmetrization for ideal bosons
(London 1938, Pathria 1972 chapter 7, Attard 2023 sections~8.2 and 8.3).
The ideal boson $\lambda$-transition,
$\rho_\mathrm{c,id} \Lambda_\mathrm{c,id}^3 = 2.612$,
can be seen in table~\ref{Tab:t1} to occur between
$T^* = 0.60$ and $T^* = 0.55$.
There is no unambiguous evidence in the table of a transition,
since the heat capacity peak is within the statistical error,
the maximum occupancy and the ground state occupancy
increase continuously
(although the former is greater than the latter
only above the ideal boson transition),
and the viscosity decreases continuously through this range.
These results are discussed individually below.

One reason for the limited evidence for the $\lambda$-transition
in the present series of simulations is its weakness
in the case of ideal bosons
(London 1938, Pathria 1972 chapter 7, Attard 2023 sections~8.2 and 8.3).
In the previous QLMC simulations for Lennard-Jones $^4$He bosons
both pure and mixed loops were used (Attard 2023 chapter~8).
Mixing ground momentum state bosons
in position loops suppresses condensation,
which makes the transition quite sharp
(Attard 2023 section~8.5).
The absence of both position loops and such mixing in the present simulations
means that the transition itself must resemble that for ideal bosons,
which prevents a divergence in the heat capacity
and precludes a sharp transition in condensation.

\subsubsection{Heat Capacity} \label{Sec:Cv}

From the classical ($S^\mathrm{perm}({\bf p}) \equiv 0$) results
in table~\ref{Tab:t1},
it can be seen that in general the kinetic energy is within
0.1--3\% of the exact result
$\beta\langle  {\cal K} \rangle^\mathrm{cl}/N = 3/2$.
The accuracy depends upon the length of the time step,
the order of the equations of motion,
and the magnitude of the thermostat parameter $\sigma^*$.
Comparing the first (not shown) and second order equations of motion,
for comparable accuracy
a ten times larger time step could be used for the latter,
which took about three time longer to evaluate per time step.
It was assumed that the parameters that were found adequate in the classical case
could be used for the quantum calculations.

The simulated heat capacity was found to be sensitive to the
value used for $\sigma$.
For example, at $T^*=0.65$
for the classical case,
at $\sigma^* = 0.2$
the heat capacity is $C_V/Nk_\mathrm{B}=1.9(4)$.
The fluctuations in kinetic energy are
$\beta^2 [\langle {\cal K}^2 \rangle - \langle {\cal K} \rangle^2 ]/N
= 1.0(2)$,
compared to the exact value of $3/2$.
At $\sigma^* = 0.5$,
the heat capacity is $C_V/Nk_\mathrm{B}=2.4(9)$,
and the fluctuations in kinetic energy are
$\beta^2 [\langle {\cal K}^2 \rangle - \langle {\cal K} \rangle^2 ]/N
= 1.3(5)$.
In the former case the average ratio
of the stochastic change in kinetic energy
at each time step to the classical adiabatic change is
$\tilde \Delta {\cal K} / \Delta^0 {\cal K}
= 0.082$, and in the latter it is $0.51$.
In the former case the viscosity is $\eta^* = 3.8(6)$,
and in the latter case 3.7(12).
From these one concludes that
the value of the stochastic parameter affects the heat capacity,
and that a value of at least $\sigma^* = 0.5$ should be used.
The larger value
has no significant effect on the viscosity.

Energy is a classically conserved variable in a microcanonical system,
and so the adiabatic energy fluctuations must be zero.
This explains why the heat capacity, as measured by
the average of the energy fluctuations, is so sensitive
to small values of the stochastic parameter $\sigma$.
(The classical adiabatic motion is the dominant contribution
to the decoherent quantum motion.)
The viscosity, on the other hand,
is measured from the fluctuations of the momentum moment,
and this is not a conserved variable in a microcanonical system.
One therefore does not expect the viscosity to be sensitive
to the value of $\sigma$.

The ideal boson heat capacity at the $\lambda$-transition is
$C_V/Nk_\mathrm{B} = 1.925$
at  $\rho_\mathrm{c,id} \Lambda_\mathrm{c,id}^3 = 2.612$. (Pathria 1972).
The greatest heat capacity in the present results
on a relatively coarse temperature grid is
$C_V/Nk_\mathrm{B} = 1.8(5)$ at $T^*=0.60$,  $\sigma^*=0.2$,
$\rho \Lambda^3 = 2.33$
or
$C_V/Nk_\mathrm{B} = 2.1(11)$
at $T^*=0.55$, $\sigma^*=1.0$, $\rho \Lambda^3 = 2.70$.
Although the statistical error is rather large,
the heat capacity in the table
for both the quantum and the classical liquid
shows a rather weak variation with temperature.
The largest quantum value is smaller than experimental values
because it is missing
position permutation loops, which diverge.
It should be larger than the ideal boson value
because it includes the potential energy contribution.
However the situation is complicated  by the fact that for interacting bosons
the ideal fugacity factor is smaller than the ideal fugacity
at the same value of $\rho \Lambda^3$,
and hence the number of condensed bosons and their fluctuations
are also smaller.
That the peak occurs close to the location of the ideal boson peak
is expected because ideal boson statistics
are used for the permutation entropy.

At $T^*=0.55$ the QSMD kinetic and potential energy
contributions to the heat capacity are about equal.
The discrepancy between $\langle \Delta {\cal H}^2 \rangle$
and
$\langle \Delta {\cal K}^2 \rangle + \langle \Delta {U}^2 \rangle$
was of the same order as the statistical error.
In general, but most notably for smaller values of $\sigma^*$,
the former expression was less than the latter,
which indicates a negative correlation between
the fluctuations in potential and kinetic energy,
as expected from energy conservation on the dominant classical adiabatic
part of the trajectory.
Because of the factorization of the momentum and the position integrals,
exact equality would be expected if the thermostat was adequate.
Since the main focus of this paper
is on dynamic rather than static properties,
and since position permutation loops have been neglected,
it was decided not to pursue more
reliable values for the heat capacity.

The kinetic energies in the quantum cases were significantly less
than the classical result.
This is an expected effect of  condensation
into low lying momentum states in the quantum system.

\subsubsection{Stochastic Dissipative Contribution}

The indirect effect of the stochastic parameter
was just discussed in the context of the heat capacity.
The direct influence of the thermostat can be gleaned from
$| \tilde \Delta {\cal K}/\Delta^0{\cal K} |$,
which
is the ratio of the stochastic to the decoherent quantum change
in kinetic energy per time step.
Using a parameter value $\sigma^* = 0.5$
gave a ratio of about 1.5 in most cases.
The ratio can be expected to scale quadratically with the parameter;
using $\sigma^* = 0.2$ reduced the ratio to 0.1--0.2,
and a value of $\sigma^* = 1.0$ increased the ratio to 2--3.
A large value of the thermostat parameter speeds equilibration,
but should otherwise have no effect on static equilibrium averages
(apart from the energy fluctuations).
From the table, it appears to have little
effect on dynamic quantities such as the viscosity.
As mentioned,
the momentum moment is not a conserved variable
in the microcanonical system.

For the quantum case at $T^*=0.55$ (below the ideal boson transition),
the value $\sigma^* = 0.2$ gives a significantly smaller
ground momentum state occupancy  than the value at $\sigma^* = 1.0$
for the same time step.
Also, the kinetic energy is significantly higher,
and the heat capacity is lower.
A smaller time step at $\sigma^* = 0.2$
increases the occupancy slightly.
Broadly similar conclusions can be drawn from the results at $T^*=0.60$.
It may be that in these cases the numerical error
due to the length of the time step
is too large to be compensated completely by the thermostat.
In general terms, the smaller the time step the more reliable are the results.

\subsubsection{Occupancy}

The quantity {\tt avocc} in table~\ref{Tab:t1}
is the total number of bosons divided
by the average number of occupied states,
which were counted with the quantized occupancy, equation~(\ref{Eq:DisctsOcc}).
One can see that for the quantum liquid {\tt avocc} is on the order of two,
and that it increases with decreasing temperature.
It is about 50\% larger in the quantum case than in the classical case.

The average maximum occupancy {\tt maxocc}
is the average occupancy of the most occupied state at each instant.
It varies between about 30 and 150
in the quantum case,
again increasing with decreasing temperature.
At $T^*=0.60$
in the classical case it is about 6
compared to about 77 in the quantum case.
These results are for $N=1000$;
for $N=5000$ in the classical case ${\tt maxocc} =7.47(7)$,
which tends to confirm
that the occupancy of a state is an intensive variable that is independent
of the system size
(Attard 2023 sections~8.2.4, 8.6.6, 9.2).
(The reason for the small increase in the maximum for the five times larger system
appears to be the finer resolution of the states in the larger system;
measuring the maximum of a curve by sampling
produces a larger value if more samples are used.)
The confirmation of the intensive nature of occupancy
is even stronger and less ambiguous for the occupancy
of the ground momentum state itself.

By definition the average maximum occupancy must be greater than or equal to
the average occupancy of the ground momentum state, $\langle N_{\bf 0} \rangle$.
Below the ideal boson condensation transition,
$\rho_\mathrm{c,id} \Lambda_\mathrm{c,id}^3 = 2.612$,
at $T^*=0.55$ ($\rho^* = 0.905$, $\rho \Lambda^3 = 2.702$)
the maximum occupancy and the ground state occupancy
are equal to within statistical error.
It is emphasized that in this case about 15\% of the bosons in the system
are condensed into the ground momentum state,
which of course means that 85\% of the bosons
are not in the ground momentum state.
The relative number of bosons in the ground momentum state
is expected to go to zero in the thermodynamic limit
(Attard 2023 sections~8.2.4, 8.6.6, 9.2).

To see this another way,
at $T^*=0.55$,
in one series of quantum runs at $\sigma^*=1.0$,
the average ground momentum state occupancy was 147(23)
and the average occupancy of a single first excited momentum state was 24(4).
Since the latter is six-fold degenerate,
this means that on average there are as many bosons
in a first excited state as in the ground momentum state:
143(9) versus 147(23).
In another series of quantum runs at the same temperature and $\sigma^*=0.2$,
$\langle N_{\bf 0} \rangle  = 80(25)$ and $\langle N_{1,0,0} \rangle  = 23(3)$,
which gives on average 136(8)  bosons in all the first excited states.
These appear to be affected by the time step,
as halving it to $\tau^*=0.5\times 10^{-4}$ with $\sigma^*=0.2$,
gives
$\langle N_{\bf 0} \rangle  = 94(21)$
(and $\langle N_{1,0,0} \rangle  = 23.5(16)$).
This is probably the most reliable value.
The runs at the smaller time step were smoother
and gave smaller statistical error for the same number of time steps
(ie.\ half the total time of the trajectory),
and in retrospect it would have been better
to have run all the quantum cases with the smaller time step.

Due to permutation entropy,
the `stickiness' of the ground momentum state
is greater than that of any one first excited momentum state.
But it must be clearly understood that these occupancy numbers,
${\cal O}(10^2)$,
will not change in taking the thermodynamic limit,
which is to say that the permutation entropy for the ground momentum state
does not diverge.

These results contradict Einstein (1924) and London (1938),
who assumed that Bose-Einstein condensation is solely into the ground state.
It is not even predominantly into the ground state,
since there are more condensed bosons \emph{not} in the ground state
than are in the ground state.
(A condensed boson is here taken to be one in a multiply occupied momentum state.)
Instead they reinforce the present author's contention
that condensation is into multiple low-lying momentum states
(Attard 2023 sections~8.2.4, 8.6.6, 9.2).

There is also a contradiction between
the present result of ${\cal O}(10^2)$ bosons
in the momentum ground state at $T^*=0.55$,
with a previous estimate of ${\cal O}(10^6)$ bosons
in  the critical velocity momentum state for capillary superfluid flow
(Attard 2023 section 9.4.3).
That number is based on the binary division approximation,
which is inconsistent in that it equates
the number of condensed bosons (extensive)
to the number in a single momentum state (intensive).
It would appear that the earlier estimate
should be reinterpreted as referring to a range of momentum states
rather than to a single critical momentum state.
This would make more sense in that superfluid flow is macroscopic,
and the total occupancy of a macroscopic number
of momentum states is macroscopic.

\begin{figure}[t]
\centerline{ \resizebox{8cm}{!}{ \includegraphics*{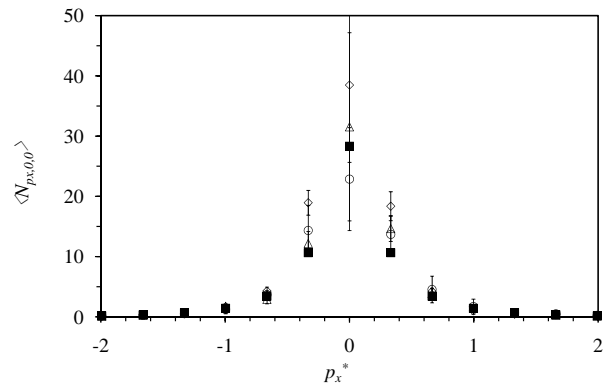} } }
\caption{\label{Fig:Npx}
Average momentum state occupancy $\langle N_{p_x,0,0} \rangle$
as a function of momentum $p_x^* = p_x /\sqrt{mk_\mathrm{B}T}$
along an axis
($T^*=0.60$, $\rho^* = 0.887$, $N=1,000$, $\sigma^*=0.2$).
The circles are for $\kappa=10$ $c=1$ Eq.~(\ref{Eq:poly})
$1.5\times10^6$ time steps,
the triangles are for $\kappa=11$ $c=1$ Eq.~(\ref{Eq:hyp})
$2\times10^6$ time steps,
and the diamonds  are for $\kappa=10$ $c=1.04$ Eq.~(\ref{Eq:poly})
$10^6$ time steps.
The solid squares are the ideal boson result,
$N(p_x,0,0) = ze^{-\beta p_x^2/2m}/[1-ze^{-\beta p_x^2/2m}]$,
using the fugacity corresponding to the ground momentum state occupancy
averaged from the three simulation results,
$z=\langle N_{\bf 0}\rangle /[1+\langle N_{\bf 0}\rangle ]$.
The error bars are twice the standard error (95\% confidence level).
}
\end{figure}

Figure~\ref{Fig:Npx} shows the average occupancy of momentum states
along an axis.
These are from a different series of simulations
to the data shown in table~\ref{Tab:t1}
and test  various umbrella sampling parameters.
Values of $\kappa$
in Eq.~(\ref{Eq:poly}) and in Eq.~(\ref{Eq:hyp})
should not be compared.
No systematic studies of the behavior with this parameter were carried out beyond
noting that values around 10 were better than values around 2.
It was concluded the hyperbolic form, Eq.~(\ref{Eq:hyp}),
was more statistically efficient than the polynomial form Eq.~(\ref{Eq:poly}).

In both case the umbrella weight ratio was brought closer to unity
by judicious choice of the parameter $c$.
Of course what matters is the fluctuations in this ratio,
but these are necessarily smaller the smaller it is.
At $T^*=0.65$, 
for $\kappa=11$ (hyperbolic umbrella fractional occupancy, Eq.~(\ref{Eq:hyp}))
using $c=1.040$ gave average umbrella weight ratio 6(6)
and using $c=1.035$ gave $6(10)\times10^3$.
The potential energy was $\beta U/N = $ -9.732(16) and -9.725(19), respectively.
The kinetic energy was $\beta {\cal K}/N = $ 0.96(4) and 0.90(2), respectively.
The {\tt maxocc} was 34(4) and 34(3).
The error estimates are for the same number of time steps.
Despite the fact that the average umbrella weight ratio
is quite sensitive to the value of the parameter $c$,
there is no strong evidence that using it
to make the ratio close to unity significantly reduces the statistical error.

The data in figure~\ref{Fig:Npx}
is at $T^*=0.6$, $\rho^* =0.8872$,
where $\rho \Lambda^3=2.33$.
This is just above the ideal boson condensation transition,
$\rho_\mathrm{c,id} \Lambda_\mathrm{c,id}^3= 2.612$.

The classical distribution is
$N(p_x,0,0) = N({\bf 0}) e^{-\beta p_x^2/2m}$ (not shown).
Compared to this
the quantum distribution in figure~\ref{Fig:Npx}
is much higher and more sharply peaked
For comparison,
the classical liquid has ground momentum state occupancy
 $N({\bf 0})=2.18(5)$ (table~\ref{Tab:t1})
compared to the quantum value of 53(8)--86(12) (table~\ref{Tab:t1})
or 22(9)--38(13) (figure~\ref{Fig:Npx}).

Perhaps the most important point of the figure
is the agreement
between the known analytic result for ideal bosons,
$N(p_x,0,0) = ze^{-\beta p_x^2/2m}/[1-ze^{-\beta p_x^2/2m}]$,
and the present simulations.
This confirms the validity of the present equations of motion,
molecular dynamics simulation algorithm,
and the continuous distribution umbrella sampling technique.
The fugacity fitted here, $z=0.966$,
is the product of the ideal and excess factors,
$z= z^\mathrm{id} z^\mathrm{ex} $,
with $z^\mathrm{id} = 0.995$
and $ z^\mathrm{ex} = VQ^\mathrm{cl}(N,V,T)/Q^\mathrm{cl}(N+1,V,T) $.
The occupation of the ground momentum state for Lennard-Jones bosons
is almost an order of magnitude smaller (28 versus 197)
than would be predicted for ideal bosons
at the same density and thermal wavelength.

The rather large statistical error for the ground momentum state,
as can be seen from the error bars in  figure~\ref{Fig:Npx}
or by comparing the results to those table~\ref{Tab:t1},
is an indicator that the fluctuations in occupancy are large and long-lived.
For ideal bosons,
the relative mean square fluctuations are close to unity
in the condensed regime,
$[\langle N_{\bf a}^2 \rangle_\mathrm{id}^+
- (\langle N_{\bf a} \rangle_\mathrm{id}^+)^2]
/(\langle N_{\bf a} \rangle_\mathrm{id}^+)^2
= 1 + 1/\langle N_{\bf a} \rangle_\mathrm{id}^+$
(Pathria 1972 equation~(6.3.9), Attard 2023 equation~(7.159)).

\begin{figure}[t]
\centerline{ \resizebox{8cm}{!}{ \includegraphics*{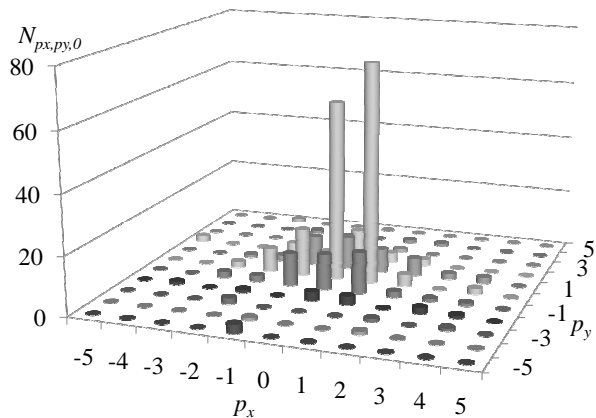} } }
\caption{\label{Fig:N(xy0)}
Instantaneous snapshot of the momentum state boson occupancy in the $p_z=0$ plane
($T^*=0.60$, $\rho^* = 0.887$, $N=1,000$, $\sigma^*=0.2$,
$\kappa=11$, $c=1.035$, Eq.~(\ref{Eq:hyp})).
}
\end{figure}

Figure~\ref{Fig:N(xy0)} gives a typical snapshot
of the instantaneous boson occupancy
of the low lying momentum states in the $p_z=0$ plane.
Of course this was taken from a trajectory
corresponding to the umbrella distribution,
and it can be regarded as typical of the actual probability distribution
only in so far as the former approximates the latter.
The envelope of the approximately Gaussian distribution
apparent in figure~\ref{Fig:Npx} can be seen.
What is also noticeable is the roughness of the instantaneous distribution,
with adjacent states often having markedly different occupancies.
In this particular snapshot, the ground momentum state
is the second most occupied state in the $p_z=0$ plane.
This is by no means atypical,
as the results for {\tt maxocc} and $\langle N_{\bf 0}\rangle$
in table~\ref{Tab:t1} confirm.
In none of the snapshots from the simulations
that have been examined for
$ \rho \Lambda^3 > \rho_\mathrm{c,id} \Lambda_\mathrm{c,id}^3  = 2.612$
is there any indication
that a significantly greater number of bosons
occupy the ground momentum state
in preference to the nearby excited states.
However, for $T^*=0.55$ and $ \rho \Lambda^3 = 2.70$,
the snapshots confirm that the ground momentum state
is almost always the most occupied momentum state,
with only two exceptions in twenty four snapshots noted.

\subsubsection{Viscosity} \label{Sec:Visco}

As a check of the present algorithm
and computer program,
a comparison was made with previous classical calculations
of the Lennard-Jones viscosity.
Previous classical non-equilibrium stochastic dissipative
molecular dynamics (NESMD) simulations for a Lennard-Jones fluid
($R_\mathrm{cut}^*=2.5$)
at $T^*=2$ and $\rho^*=0.8$ gave $\eta^*=1.88(3)$ (Attard 2018a).
At the same state point the present QSMD program
in classical mode
(ie.\ $S^\mathrm{perm}({\bf p}) \equiv 0$
and
$\nabla_p S^\mathrm{perm}({\bf p}) \equiv {\bf 0}$),
with $R_\mathrm{cut}^*=3.5$ gave $\eta^*=2.16(15)$.
Since the non-equilibrium algorithm is different
to the equilibrium Green-Kubo approach,
and since the two programs used to calculate the viscosity
are unrelated to each other
(except for the neighbor table subroutines),
this confirms that the present program is free from error,
at least in the classical parts of the computation.

\begin{figure}[t!]
\centerline{ \resizebox{7.8cm}{!}{ \includegraphics*{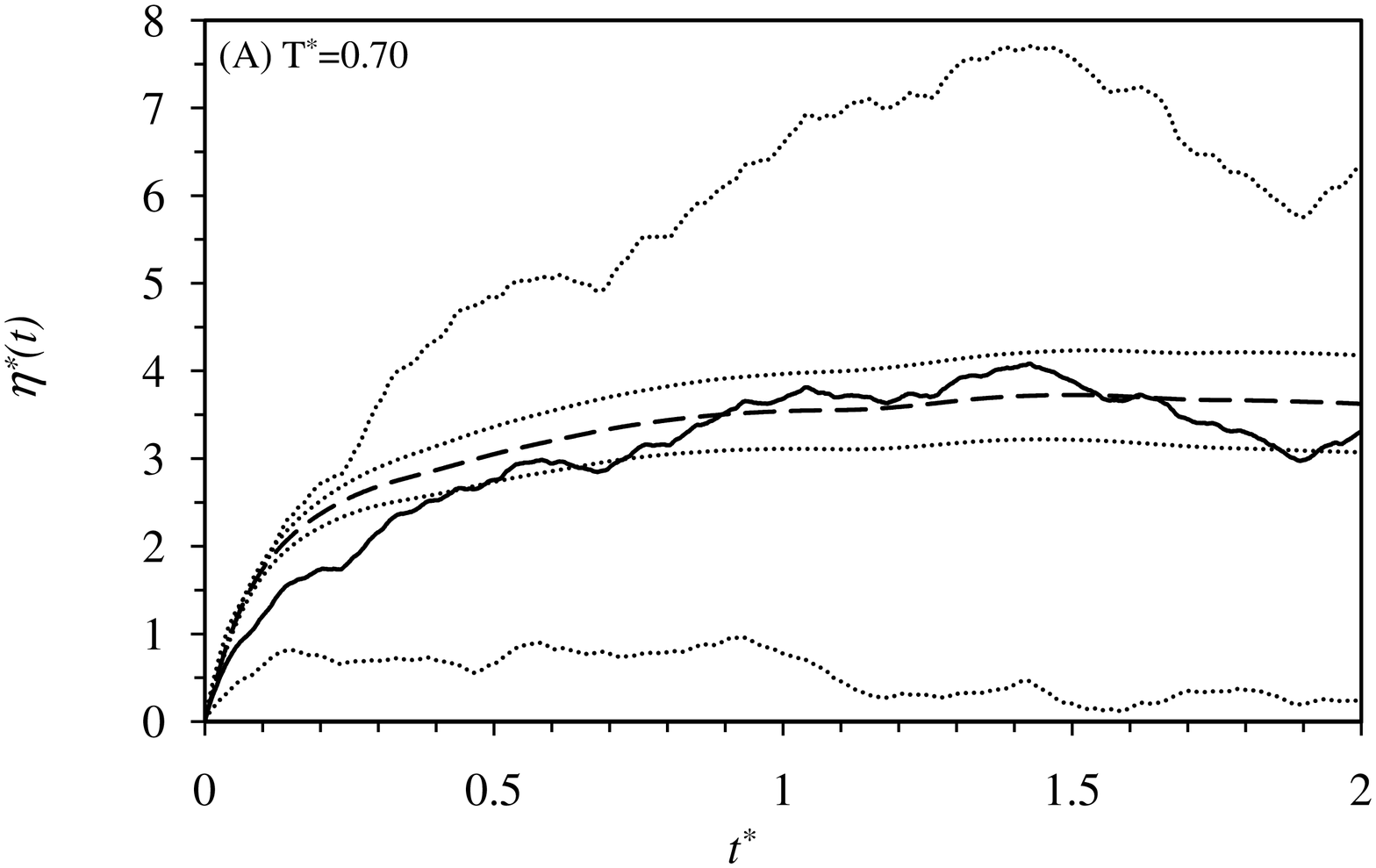} } }
\centerline{ \resizebox{7.8cm}{!}{ \includegraphics*{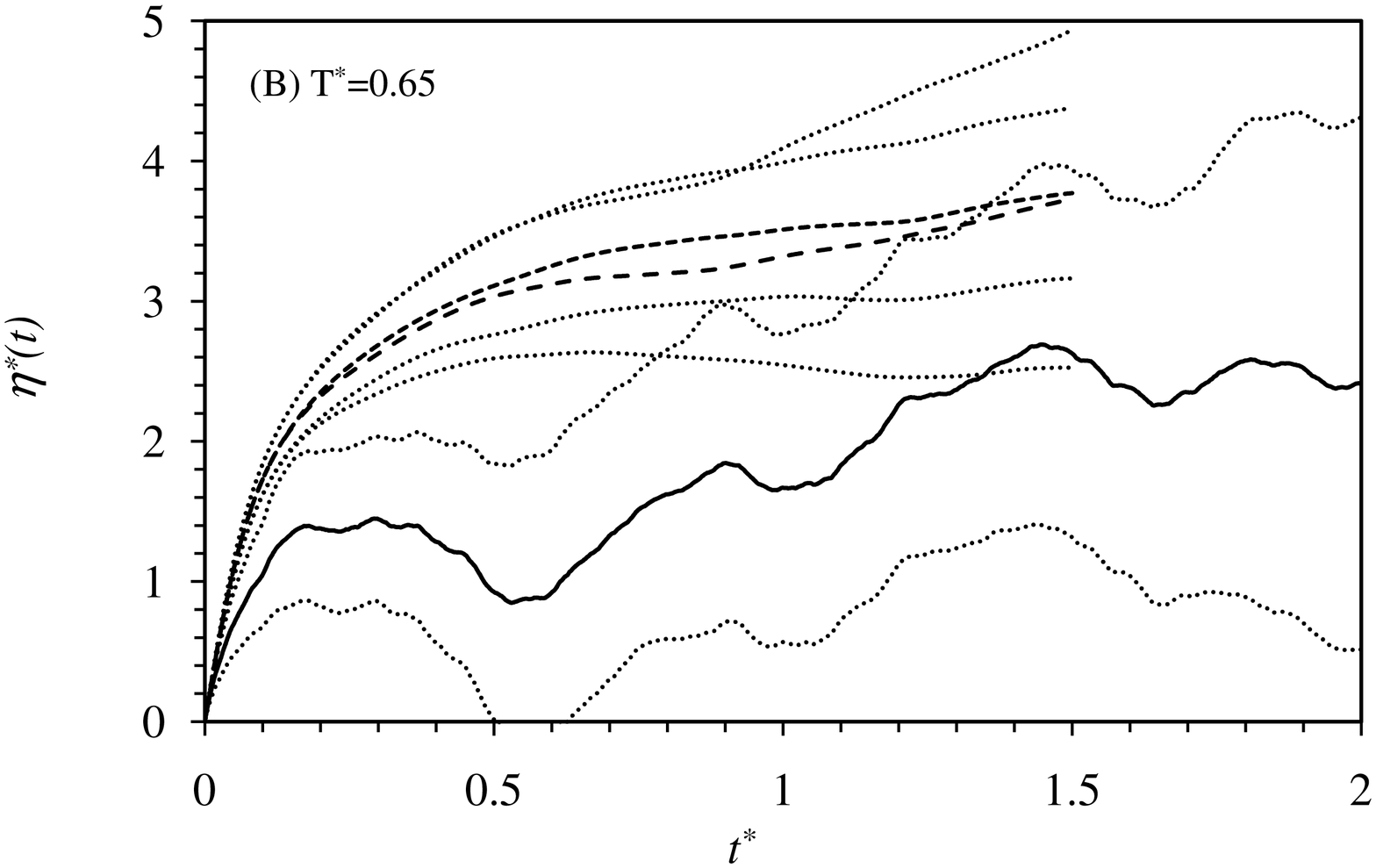} } }
\centerline{ \resizebox{7.8cm}{!}{ \includegraphics*{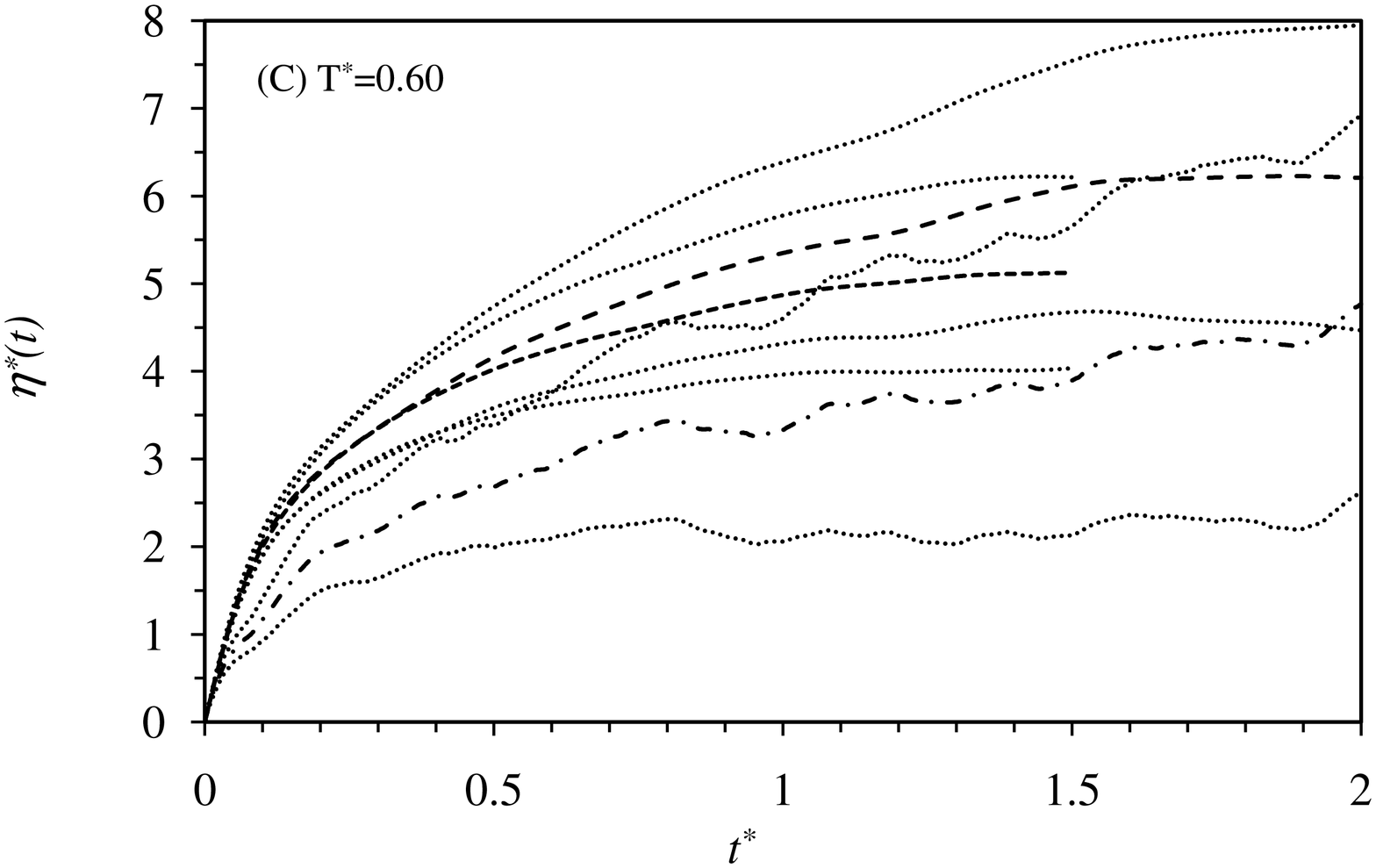} } }
\centerline{ \resizebox{7.8cm}{!}{ \includegraphics*{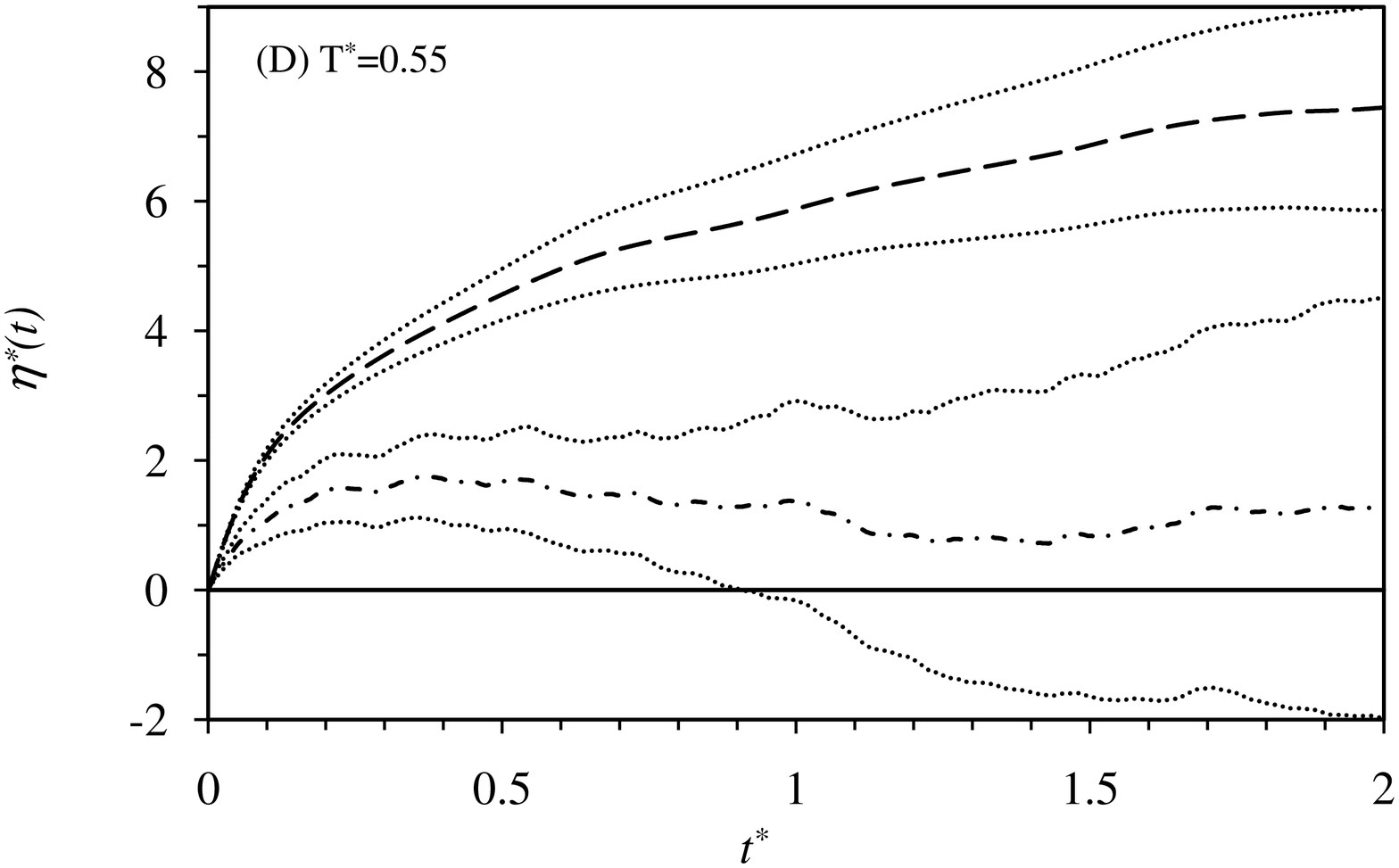} } }
\caption{\label{Fig:visc}
The viscosity 
for the quantum (solid and dash-dot curves)
and the classical (dash curves) saturated liquids.
The time step is $\tau^* = 10^{-4}$
except for the dash-dot curves where $\tau^* = 0.5 \times 10^{-4}$.
The stochastic parameter is $\sigma^*=1.0$ (solid and long-dash curves),
$\sigma^*=0.5$ (medium-dash curves),
or $\sigma^*=0.2$ (dash-dot and short-dash curves).
The dotted curves give the 95\% confidence interval.
}
\end{figure}

Figure~\ref{Fig:visc}  shows the viscosity time correlation function,
equation~(\ref{Eq:eta(t)}),
averaged over all six off-diagonal matrix elements.
The viscosity must go to zero at $t=0$,
as can be seen directly from equation~(\ref{Eq:eta(t)}),
or from the usual parity arguments based on time reversibility
(Attard 2012 equation~(2.60)).
It can be seen that the classical viscosity rises rapidly
to a rather broad maximum.
The maximum value is `the' viscosity.
The curvature of the plateau region decreases with increasing system size,
and it can be expected to become flat in the thermodynamic limit.
In so far as hydrodynamics usually refers to steady state
or slowly varying flows,
the long-time limit of the shear viscosity time correlation function
is the appropriate value,
which is likely the maximum value for a finite-sized system
when the thermodynamic limit is taken.
In the present cases the maximum in the classical viscosity
occurs for $t^*\alt 2.0$.
Because the curves are flat, missing the location of the maximum does not greatly
affect the value estimated for the viscosity.

The classical viscosities at $T^*=0.65$ and $T^*=0.60$
are insensitive to the two different values of the stochastic parameter $\sigma^*$
that were used.
Across the temperature range, it can be seen that
the classical viscosity increases with decreasing temperature
on the saturation curve.

The quantum viscosities are more noisy
than their classical counterparts
despite the fact they they were generally run for longer.
No doubt that this is due at least in part to the large fluctuations
in the momentum state occupancy that occur in the quantum case.
It is also likely due to the umbrella sampling.

There are two contributions
to the difference between the viscosities of the quantum and classical liquids.
First is the quantum versus classical statistics,
which amongst other things enhances the multiple occupancy
of low-lying momentum states in the quantum case.
Second is the decoherent quantum term,
$\dot q^{0,\mathrm{qu}}_{iz}  p_{xi}
= - T p_{xi} \nabla_{p,iz} S^\mathrm{perm}({\bf p})$,
in the rate of change of the first momentum moment,
equation~(\ref{Eq:.P0zx}).
If this 
is set to zero,
then the viscosity of the quantum liquid generally increases,
and at lower temperatures it can actually become larger than the viscosity
of the classical liquid. 

The quantum viscosity
is obtained by integrating the time correlation function
over the thermostatted, umbrella trajectory
(ie.\ the one that invokes the continuous occupancy).
There are two approximations here:
First, and likely less serious, is using the thermostatted rather than
the decoherent quantum trajectory.
Second, is using the gradient of the continuous occupancy
for the transitions at each step on the trajectory.
No correction was made by invoking the product of the ratios
of the exact to the umbrella transition probability along the trajectory.
(Because the exact gradient of the permutation entropy
would be required for the exact transition probability,
and if it were numerically feasible to compute this
there would be no need to use continuous occupancy or umbrella sampling.)

Despite the noise in the quantum cases
one can still draw some conclusions by comparing
each with the corresponding classical case.
At the highest temperature, $T^*=0.70$,
the quantum viscosity more or less coincides with the classical viscosity.
It should be noted that even at this temperature,
which is well above the ideal boson $\lambda$-transition
($\rho \Lambda^3= 1.76$ compared to
$\rho_\mathrm{c,id} \Lambda_\mathrm{c,id}^3= 2.612$),
the present QSMD simulations,
which neglect position permutation loops
but retain pure momentum permutation loops,
give a much higher occupancy for the ground momentum  state
in the quantum than in the classical case,
(26(4) compared to 1.75(3), table~\ref{Tab:t1}).
The occupancy of other low lying momentum states is similarly enhanced.
This no doubt contributes in part
to the reduced viscosity on small time intervals,
which might be interpreted as evidence for superfluidity
arising from a dilute mixture of condensed and uncondensed bosons.

QLMC simulations show that including mixed position and momentum
permutation loops suppresses condensation into the ground momentum  state
above the $\lambda$-transition temperature (Attard 2023 section~8.4).
This appears to be the reason that the superfluid transition
measured experimentally coincides with the $\lambda$-transition.
These mixed loops are neglected in the present QSMD simulations,
as are all position permutation loops.
This explains why here there is some condensation into the ground
and other low lying momentum states
above the ideal boson $\lambda$-transition temperature,
and the (apparently consequent) reduction in the viscosity.

At $T^*=0.65$,
the quantum viscosity lies significantly below the classical velocity
on short time intervals
and at longer time scales it appears to level off below the classical value.
The occupancy of the ground momentum state is 45(12) in the quantum case
and 1.99(4) in the classical case (table~\ref{Tab:t1}).
The occupancy of other low lying momentum states is also enhanced.
At $T^*=0.60$,
the quantum viscosity lies significantly below the classical velocity
for the entire time interval shown.
Toward the end of the time interval it appears to be rising
or perhaps to be leveling off,
but the statistical error is rather too large to be certain about this.
Whereas at $T^*=0.65$ the average maximum occupancy
was greater than the occupancy of the ground momentum state
(46(9) compared to 37(9) for $\tau^*=10^{-4}$),
at $T^*=0.60$  the two are about the same,
(86(11) compared to 86(12) for $\tau^*=10^{-4}$;
63(5) compared to 52(8) for  for $\tau^*=5\times 10^{-5}$).
This says that ground momentum state occupancy
is not the only contributing factor to the reduction in viscosity.

At $T^*=0.55$,
the quantum viscosity lies significantly below the classical viscosity
over the entire time interval exhibited.
In fact, for $t^* \agt 1$ the viscosity is zero
within the statistical error.
This probably should not be taken too literally;
the liquid contains a significant fraction of bosons
in few occupied momentum states,
and one would expect the interactions amongst these to lead
to a non-zero viscosity.
It is not clear what will happen for time intervals larger than those studied,
but one might guess that the quantum viscosity levels off
to its value at the end of the short time interval.

In summary one can conclude
that at low temperatures the viscosity of the quantum liquid is less
than that of the classical liquid,
and the size of the decrement increases with decreasing temperature.
There is no sharp transition in viscosity with temperature
in the present QSMD simulations
because they neglect position permutation loops and mixed permutation loops,
which suppress boson condensation above the $\lambda$-transition
(Attard 2023 section~8.4).
Also, the Lennard-Jones pair potential
contributes to the fugacity so that the condensation
for the present interacting bosons is quantitatively different
to that for ideal bosons.

\subsubsection{Trajectory} \label{Sec:Traj}

\begin{figure}[t]
\centerline{ \resizebox{8cm}{!}{ \includegraphics*{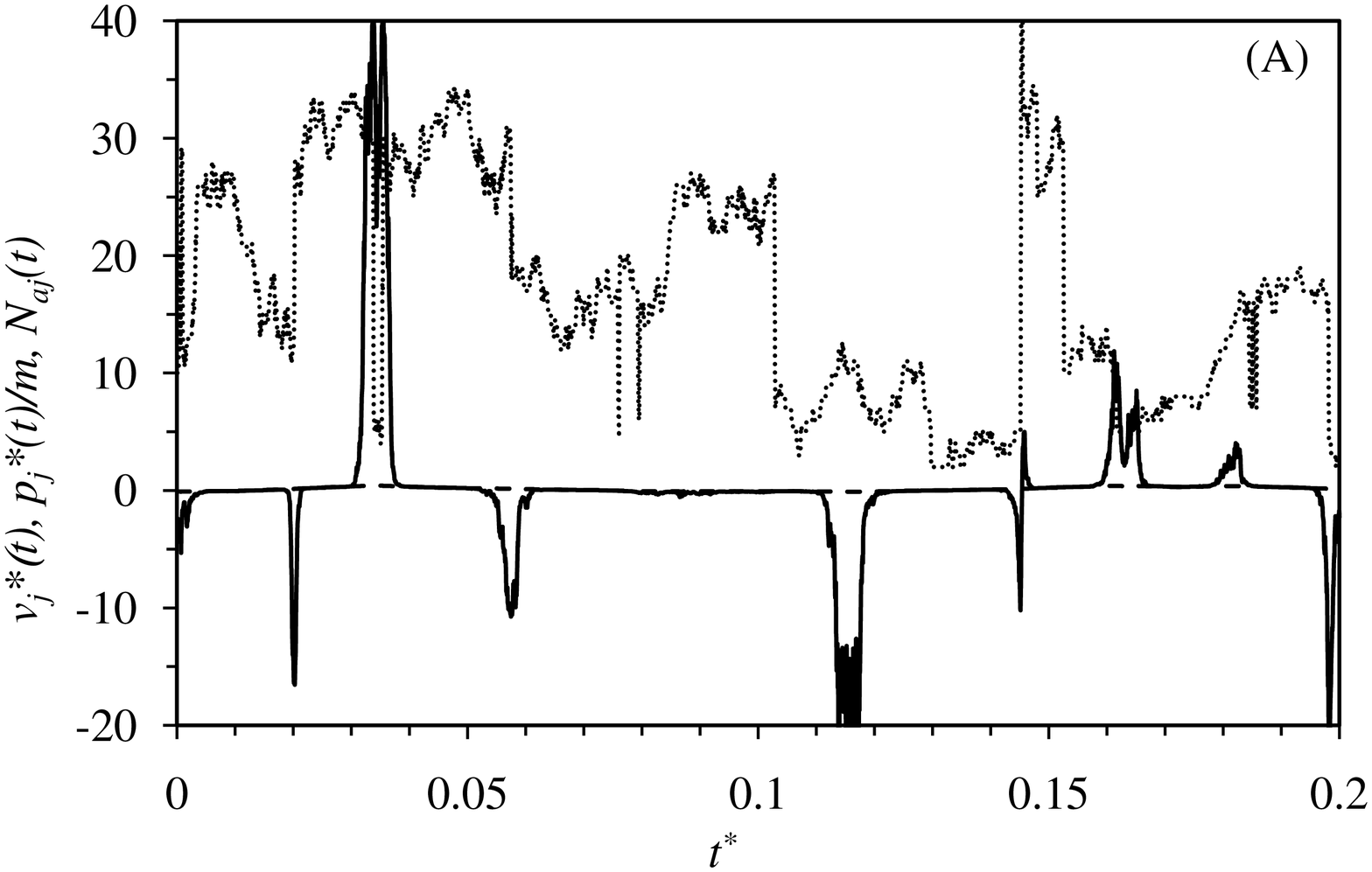} } }
\centerline{ \resizebox{8cm}{!}{ \includegraphics*{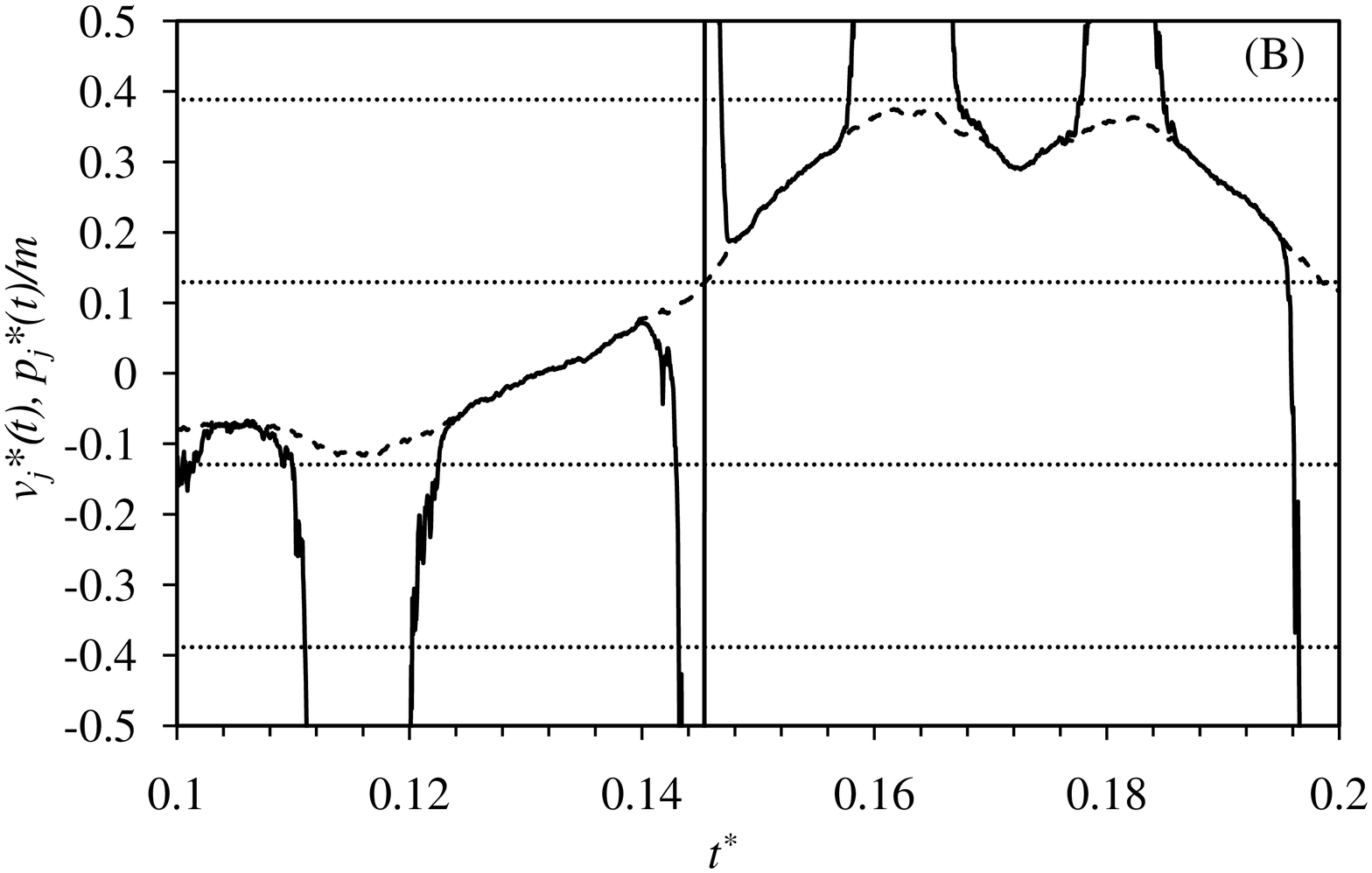} } }
\caption{\label{Fig:traj}
The $x$-components of the trajectory of a condensed boson in phase space
($T^*=0.55$, $\tau^* = 5 \times 10^{-5}$, $\sigma^*=0.2$,
$\kappa=11$ in equation~(\ref{Eq:hyp})).
The solid curve is the velocity
(ie.\ rate of change of position),
the dashed curve is the momentum divided by mass
(mostly obscured),
and the dotted curve is the discrete occupancy of the momentum state of the boson.
(B) is a magnification of a portion of (A),
with the dotted lines marking the boundaries of the momentum states.
}
\end{figure}

Figure~\ref{Fig:traj}
shows a randomly chosen trajectory of one of the bosons in the system
at $T^*=0.55$, which is below the ideal boson $\lambda$-transition.
The evolution of the system is governed by the QSMD equations of motion
with continuous occupancy, equation~(\ref{Eq:hyp}) with $\kappa=11$.
The occupancy of the states visited by this boson
is typically 5--20 on this portion of its trajectory,
and it can therefore be called a condensed boson.
Taking account as well of the $y$- and $z$-components of the momentum (not shown),
for perhaps one quarter of the trajectory shown
this condensed boson is in the ground momentum  state.
(The ground momentum state of the $x$-component
is not necessarily the ground momentum state of the boson,
as it may be in an excited momentum state
of one or both of the other components.)

The occupancy shown in figure~\ref{Fig:traj}A is the discrete one.
It would be difficult to discern the continuous occupancy from
this on the scale of the figure.
Compared to the exact discrete occupancy,
the root mean square error of the continuous occupancy is 4.8\%
over this trajectory.

In this and the following figures,
the velocity of the particle, defined as the change in position
at each time step divided by the length of the time step,
is compared to the momentum divided by the mass.
In classical mechanics these would be equal.
What is noticeable in the figure
are the large spikes in velocity.
These result from jumps in the position
that are much larger than the prior classical changes.
The spikes are more or less continuous,
with each lasting on the order of $10^2$  time steps
(depending on the width),
and comprising a sequence of moves in the same direction that first increases
and then decreases in magnitude.
(Usually the velocity during the spike does not change sign,
in which case there is a nett jump away
from the original trajectory.)
Presumably these continuous spikes
result from the continuous occupancy formulation,
and in the limit $\kappa \rightarrow \infty$
they would become $\delta$-functions
(appendix~\ref{Sec:DiscOccTraj}).

Although the velocity spikes are clearly associated with the proximity
to the boundary of a momentum state (figure~\ref{Fig:traj}B),
the occupancy of the current state changes more frequently
(figure~\ref{Fig:traj}A).
This is for two reasons:
First the spike often precludes a change in momentum state
(figure~\ref{Fig:traj}B).
And second, the changes in occupancy of a highly occupied state
are most probably due
to other bosons entering or exiting the momentum state.

A portion of the trajectory is shown in detail in figure~\ref{Fig:traj}B.
There is a deal of noise apparent in the velocity spikes on this scale,
which perhaps indicates a smaller time step,
or a smaller value of $\kappa$,
would have been efficacious.
(The continuous occupancy using equation~(\ref{Eq:poly})
with  $\kappa=10$ gives roughly similar behavior,
with the spikes perhaps being smaller in length
but broader in width at the base.
From the same starting position in phase space,
the two trajectories visibly diverge after about $t^*=0.02$.)
One sees that the spikes occur when the momentum approaches the boundary
of a momentum state.
In most, but not all, cases the rate of change of momentum
reverses sign during the spike,
which means that the jump is along a position path
such that the force on the particle passes through zero
and reverses sign.
In consequence, the particle either remains in,
or returns to, the original momentum state.
Nevertheless, it is possible to escape the ground momentum state,
and the fluctuations in the occupancy of the state that the boson
is currently in are quite large (figure~\ref{Fig:traj}A).
Another point to be noticed in figure~\ref{Fig:traj}B
is that the sign of the position jump
usually has the opposite sign to the product of the current force
and the difference in occupancies of the proposed less the current momentum state,
assuming the most occupied states are those with less momentum,
(cf.\ equation~(\ref{Eq:dot-q_ja^0qu})).

In figure~\ref{Fig:traj}B
at around about $t^*=0.14$ there is a spike
during which the velocity changes sign.
Perhaps this is better identified as two separate spikes
of opposite sign.
In any case the sequence signifies a jump away from the original position
followed by a jump back to it.
This coincides with going from a few occupied to a highly occupied state,
as can be seen in figure~\ref{Fig:traj}A.
This type of jump with velocity reversal in a single spike
appears to comparatively rare for condensed bosons,
but more common for uncondensed bosons (see below).

It is overly simplistic to interpret every position jump
as a pair collision.
They are collisions only in the sense that they occur upon approach
to the boundary of a momentum state,
and a boson can only approach such a boundary if it experiences a nett force,
which of course must come from molecular interactions.

\begin{figure}[t]
\centerline{ \resizebox{8cm}{!}{ \includegraphics*{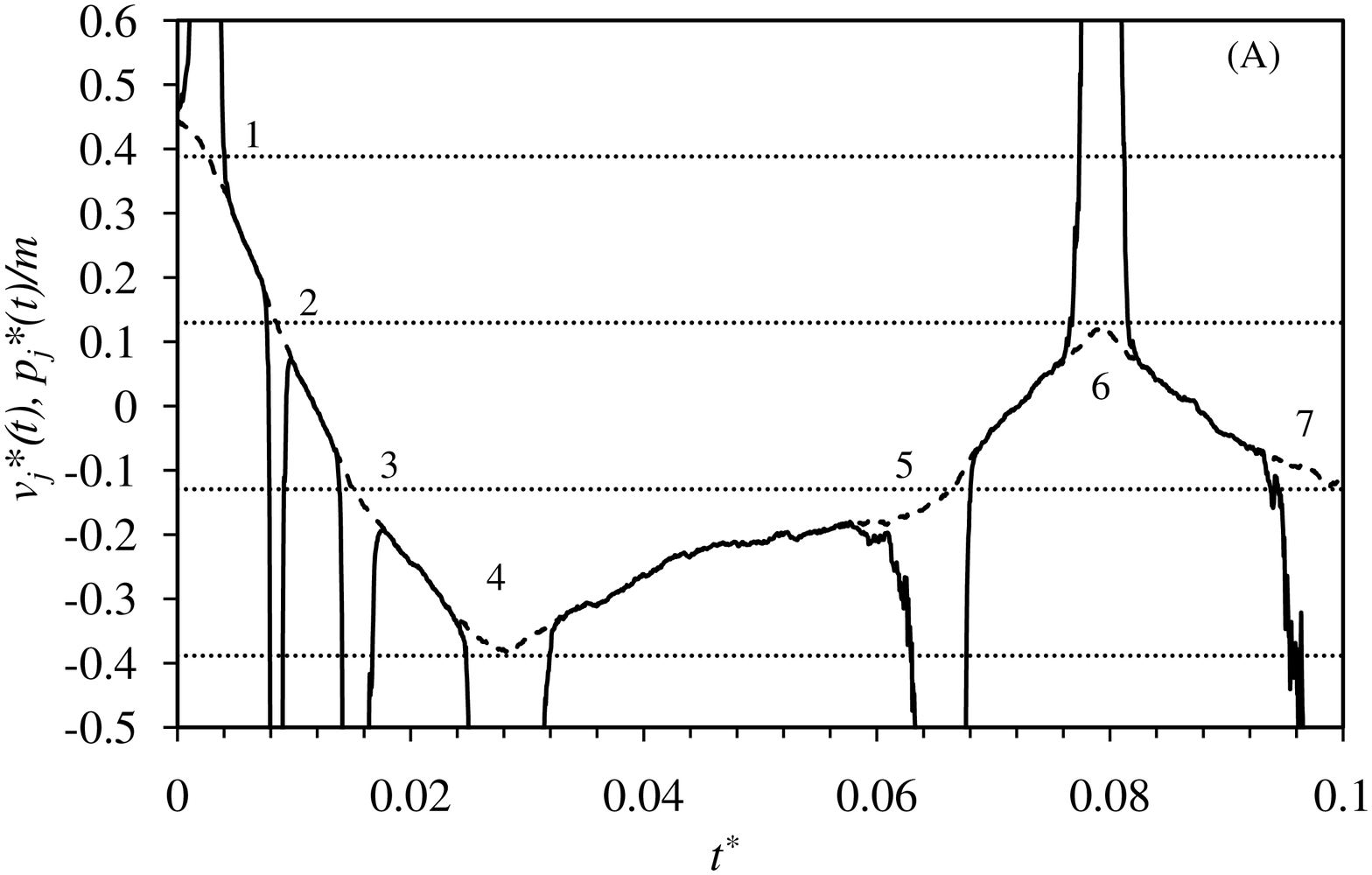} } }
\centerline{ \resizebox{8cm}{!}{ \includegraphics*{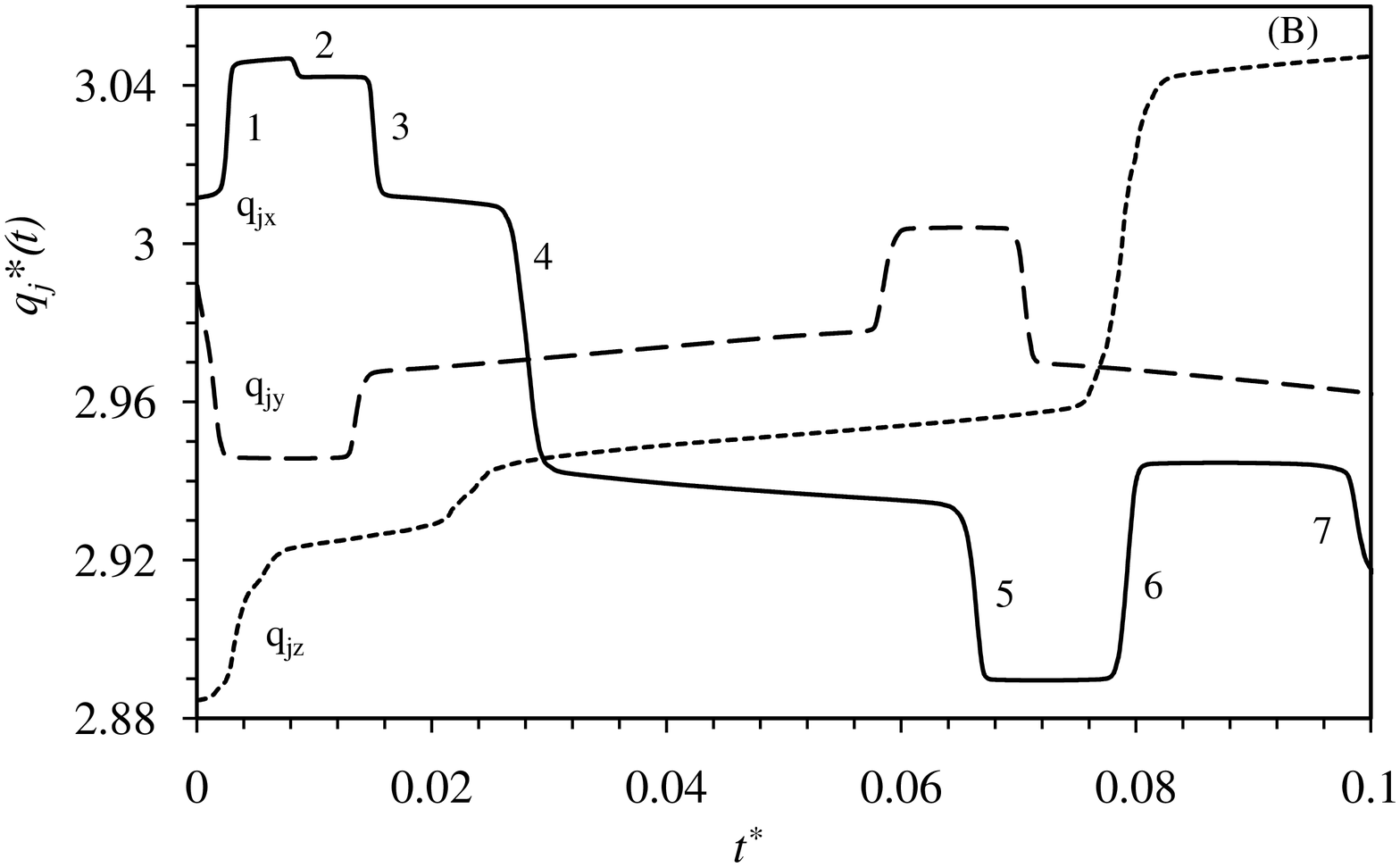} } }
\caption{\label{Fig:traj2}
(A) The $x$-components of the trajectory of a boson in phase space
(state, parameters, and curves as in figure~\ref{Fig:traj}).
(B) The components of the actual position over time,
with the numbers giving the corresponding jumps.}
\end{figure}

Figure~\ref{Fig:traj2} shows
the $x$-components of the trajectory of a boson from a different run
at the same temperature.
The first excited negative $x$-momentum state shown has an occupancy
of less than 5, as does the state prior to the first peak;
the remaining ground and first positive excited $x$-momentum states
have an occupancy of 10--30.
The trajectory in figure~\ref{Fig:traj2}A
shows spikes
near the boundaries of the momentum states.
The spikes do not always preclude a change of state,
but their sign does appear to be consistent
with the discrete occupancy, equation~(\ref{Eq:dot-q_ja^0qu}).
Spikes 4, 6, and 7 give a position jump to zero force,
which prevents a change in the momentum state.

Figure~\ref{Fig:traj2}B shows the components of the actual position
from which the velocity in figure~\ref{Fig:traj2}A is derived.
(The $y$- and $z$-components of position have been shifted
to display them on the same scale as the $x$-component.)
It can be seen that the curves are relatively smooth.
The jumps are actually quite small on the scale of the Lennard-Jones
$^4$He diameter $\sigma_\mathrm{LJ}$, which is used as the unit of length.
The length of the jump is more or less given by the width of the spike in velocity.
It can be seen that in two cases (1--3 and 5--6),
the $x$-components of the jumps come in pairs that cancel
so that the trajectory after the second jump
is more or less the continuation of it prior to the first jump.
These jumps occur close to two similar paired jumps in the $y$-component
of the position trajectory.
Since it is the boundary of the momentum state for each component
that determines the jump,
the extent  of correlation between them
would be expected to be determined
by the magnitude of the nett force on the boson.
From the molecular point of view the return jump can be understood by noting
that the first jump leaves a cavity at the original position,
and the subsequent mechanical forces act on the boson to refill it,
thereby assuaging nature's abhorrence.
The jumps in position in figure~\ref{Fig:traj2}B are quite small,
and it is not clear that a detailed molecular interpretation is called for.

\begin{figure}[t]
\centerline{ \resizebox{8cm}{!}{ \includegraphics*{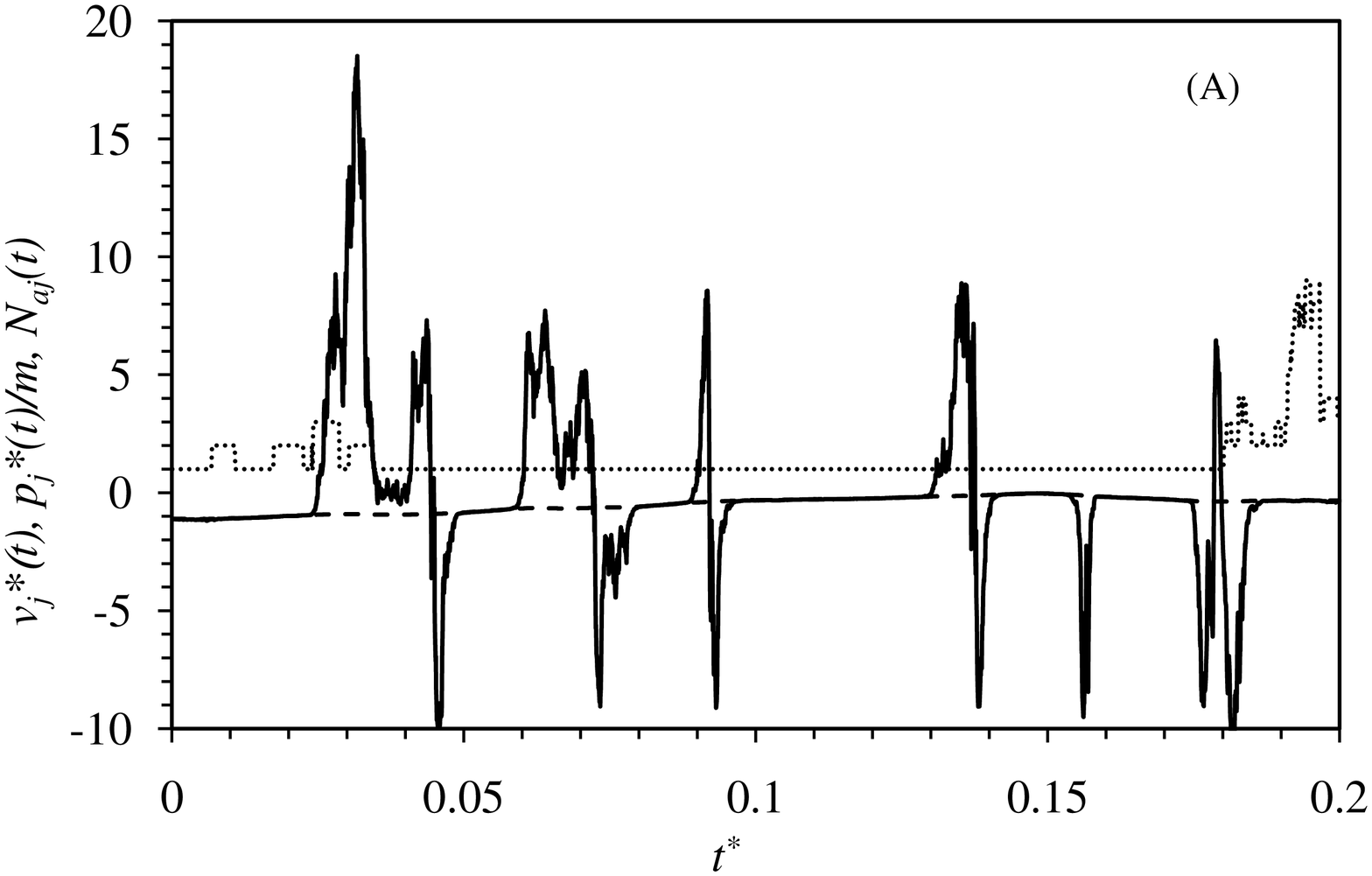} } }
\centerline{ \resizebox{8cm}{!}{ \includegraphics*{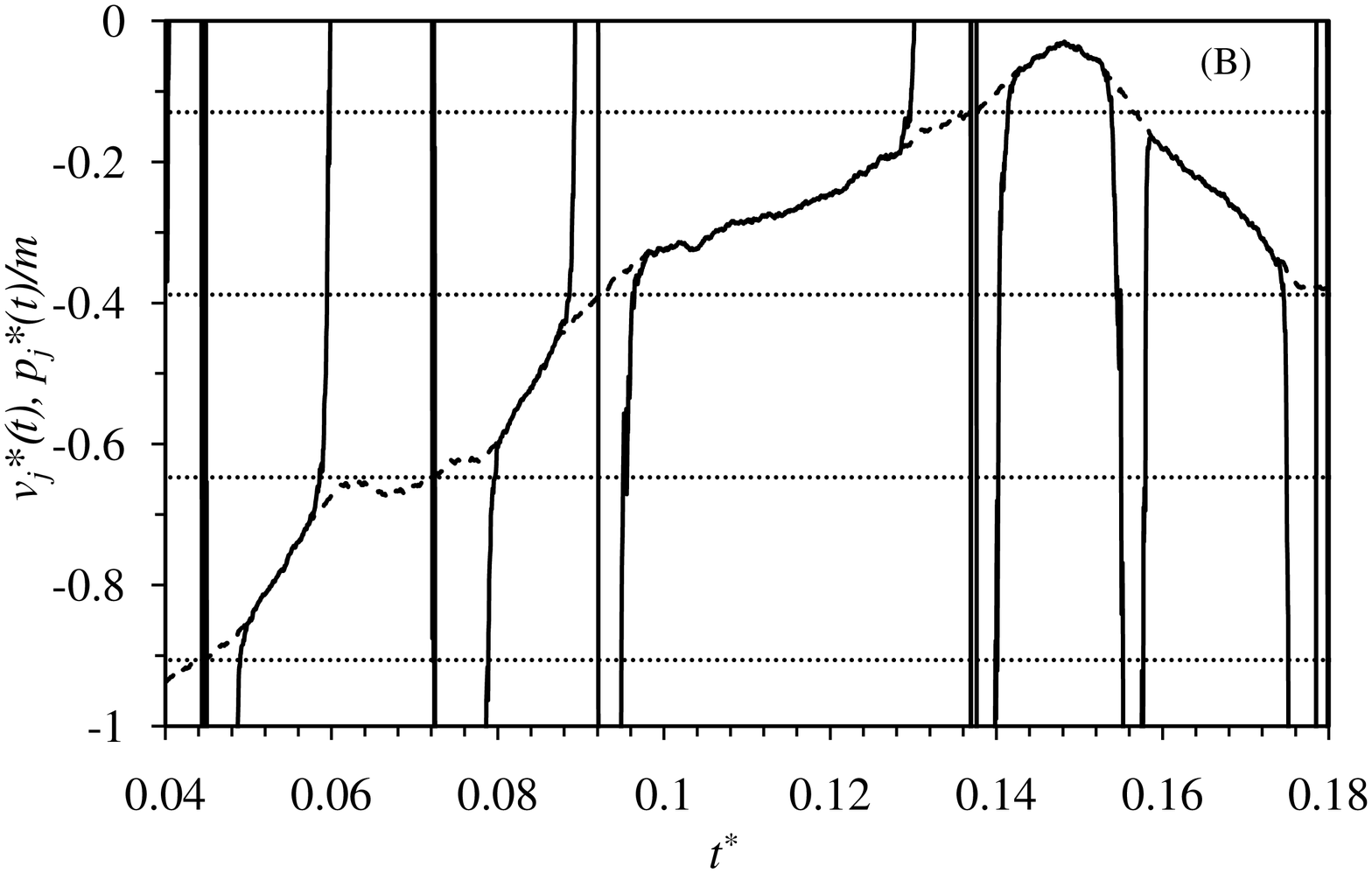} } }
\caption{\label{Fig:traj3}
The $z$-component of the trajectory of an uncondensed boson in phase space
(state, parameters, and curves as in figure~\ref{Fig:traj}).
(B) shows a detail of the trajectory during which the boson is only ever
in singly occupied momentum states.
}
\end{figure}

Figure~\ref{Fig:traj3} shows the trajectory of an uncondensed boson.
In this case the occupancy of the momentum states
that it visits is mainly 1--2,
except toward the end of the trajectory shown.
One can see that even for such an uncondensed boson
there are spikes in the velocity,
which are associated with approaching the boundary of a new momentum state
(figure~\ref{Fig:traj3}B).
The spikes are narrower and have smaller magnitudes than for condensed bosons,
which means that the jumps in position are smaller.
Many spikes show a reversal of velocity,
which effectively cancels the jump.
In figure~\ref{Fig:traj3}B
the transitions are from a singly occupied to an empty momentum state.
For most of this portion of the trajectory,
the momentum is increasing, which means that the net force
on the uncondensed boson is non-zero and positive.
The position jumps more or less cancel each other
during the velocity-reversal spikes
and so the trajectory continues on as before;
if one overlooks the spikes
the trajectory is effectively classical, adiabatic, and continuous.

This is an important observation
because the classical limit is the limit where all momentum states
are either singly occupied or else empty.
Hence an uncondensed boson such as this should follow a classical trajectory.
It is reasonable to suppose that
in the discrete occupancy limit, $\kappa \rightarrow \infty$,
the two spikes that comprise a velocity-reversal
become two equal, opposite, and superimposed $\delta$-functions,
the cancelation of which leaves a pure classical trajectory.
In appendix~\ref{Sec:DiscOccTraj}
it is shown that for discrete occupancy
there are no spikes or jumps upon the transition between momentum states
for uncondensed bosons.

\begin{figure}[t]
\centerline{ \resizebox{8cm}{!}{ \includegraphics*{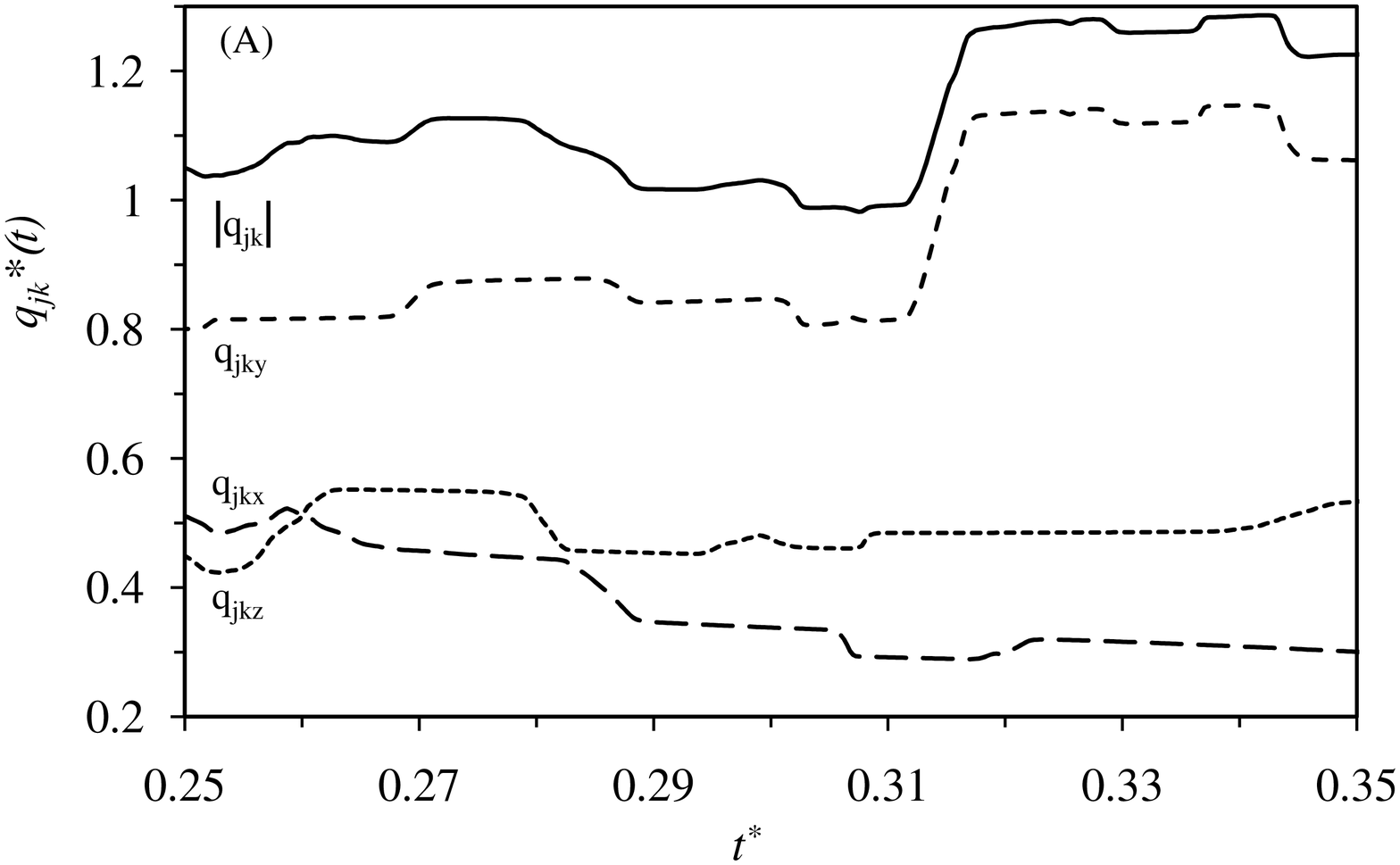} } }
\centerline{ \resizebox{8cm}{!}{ \includegraphics*{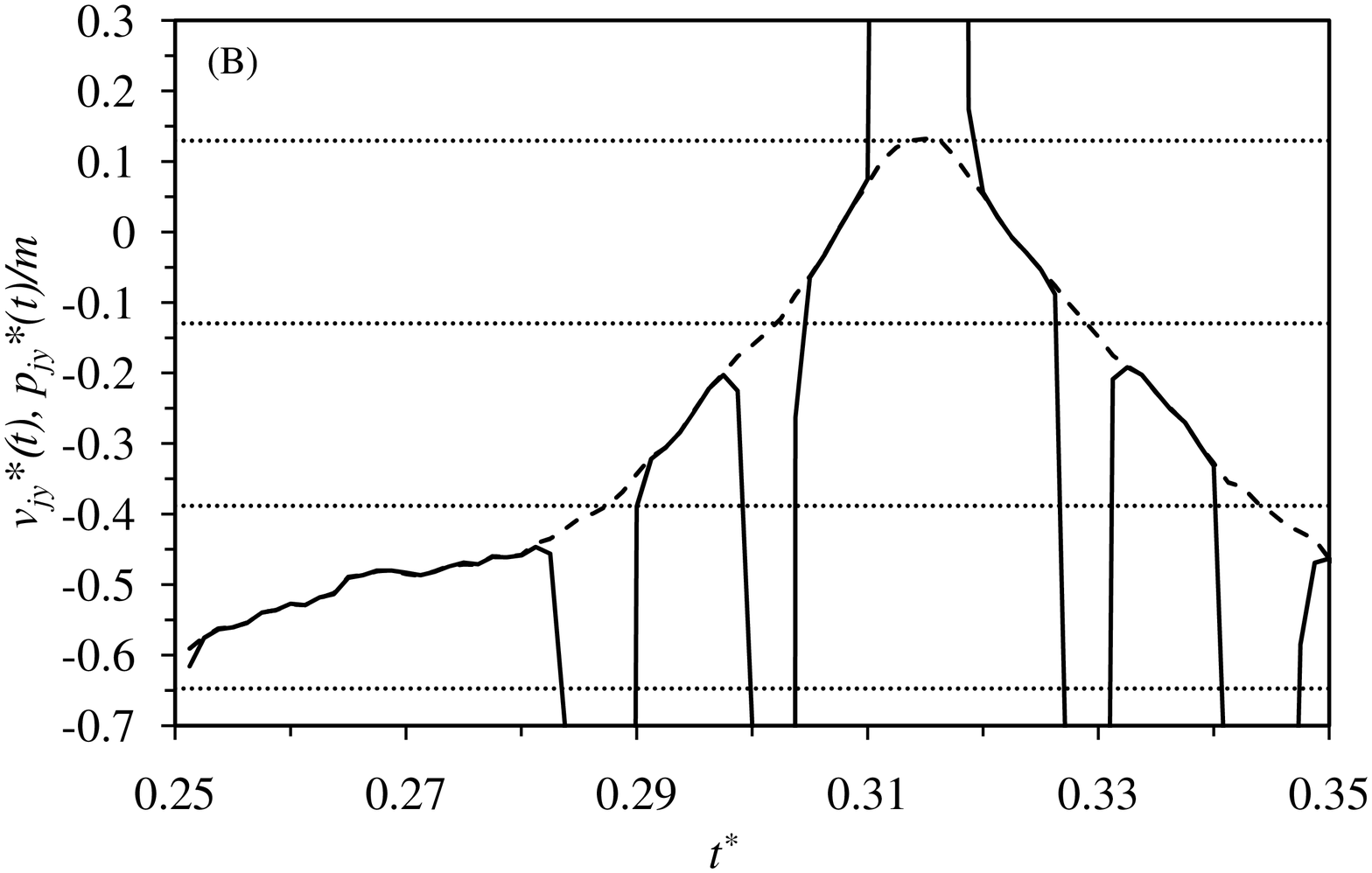} } }
\caption{\label{Fig:traj4}
Collision of a condensed boson $j$ and an uncondensed boson $k$
(state and parameters as in figure~\ref{Fig:traj}).
(A) Components of the separation, ${\bf q}_{jk} = {\bf q}_{j}-{\bf q}_{k}$.
(B) $y$-component of the velocity (solid curve)
and the momentum divided by mass (dashed curve)
of the condensed boson $j$.
}
\end{figure}

Figure~\ref{Fig:traj4} shows a collision between
a condensed boson
($j=816$, in a state with occupancy $\approx 35$ at $t^* \approx 0.31$)
and an uncondensed boson ($k=796$, occupancy $=1$).
The condensed boson was in a low-lying momentum state occupied by about 40 bosons,
but it was not the ground momentum state.
Although the trajectory is affected by the interaction potentials
with other bosons in the vicinity,
since the Lennard-Jones pair potential becomes steeply repulsive
for separations $q_{jk}^* \le 2^{1/6} = 1.12$,
one can say that the jump at $t^* \approx 0.31$
is in response to the pair collision between the two bosons.
The  $z$-separation is small and constant over the time period shown,
the $y$-separation is larger and shows a jump,
and the $x$-separation is small and decreasing.
Hence one can say that the collision occurs in the $xy$-plane,
with the relative motion in the $x$-direction
with a lateral offset (ie.\ impact parameter)  in the $y$-direction.
Figure~\ref{Fig:traj4}B shows that the collision at $t^* \approx 0.31$
does not change the $y$-momentum state of the condensed boson.
This is due to the jump  in that direction
that gives a larger separation that passes through zero force,
turning it from repulsive to attractive.
A schematic based on this `side-step' collision
is given in subsection~\ref{Sec:Landau} below,
where it is used to explain the molecular basis of superfluidity.

Figure~\ref{Fig:traj4} was obtained using equation~(\ref{Eq:hyp}) with $\kappa=11$.
The run was repeated from the same starting point
with the same sequence of pseudo-random numbers
using equation~(\ref{Eq:poly}) with $\kappa=10$.
The two trajectories for the same pair of bosons
diverged noticeably for $t^*\agt 0.02$. 
Using a different pair of bosons ($j=841$ and $k=517$,
which of all pairs had the smallest separation
one fifth of the way through the run),
which also happened to be a condensed/uncondensed pair
(the condensed boson was in a low-lying momentum state occupied by about 30 bosons,
but it was not the ground momentum state),
it was found that a qualitatively similar side-step collision occurred
(results not shown).  
This shows that the condensed boson side-step collision mechanism
on the decoherent quantum trajectory is rather common,
and that it is quite robust
and independent of the specific continuous occupancy representation.

The above trajectories can be understood
from the conservation of entropy
by the decoherent quantum equations of motion.
This can be rearranged as
\begin{eqnarray}
\dot S^0 & = &
\dot{\bf \Gamma}^0 \cdot \nabla S
\nonumber \\ & = &
-T \nabla_p [S^\mathrm{r} + S^\mathrm{perm}]
\cdot \nabla_q [S^\mathrm{r} + S^\mathrm{perm}]
\nonumber \\ & & \mbox{ }
+ T \nabla_q [S^\mathrm{r} + S^\mathrm{perm}]
\cdot \nabla_p [S^\mathrm{r} + S^\mathrm{perm}]
\nonumber \\ & = &
\big\{ -T \nabla_p S^\mathrm{r} \cdot \nabla_q S^\mathrm{r}
+ T \nabla_q S^\mathrm{r} \cdot \nabla_p S^\mathrm{r} \big\}
\nonumber \\ & & \mbox{ }
+ \big\{  -T \nabla_p   S^\mathrm{perm}({\bf p}) \cdot \nabla_q S^\mathrm{r}
\nonumber \\ & & \mbox{ }
+ T \nabla_q S^\mathrm{r} \cdot \nabla_p   S^\mathrm{perm}({\bf p}) \big\}
\nonumber \\ & = &
\frac{-1}{T} \dot{\bf q}^\mathrm{0,qu} \cdot \nabla_q U({\bf q})
+ \dot{\bf p}^\mathrm{0,cl} \cdot \nabla_p   S^\mathrm{perm}({\bf p}) .
\end{eqnarray}
This is  obviously zero already at the second equality.
But it is interesting to separate out the classical adiabatic
change of reservoir entropy in the first set of braces in the third equality,
which itself is zero,
and to rewrite the second set of braces
as the sum of the quantum change in potential energy
and the classical change in permutation entropy.
This shows that the quantum spike, $\dot{\bf q}^\mathrm{0,qu} $,
gives a jump rate in position that
changes the potential energy and hence the reservoir entropy
to exactly cancel the change in permutation entropy
that is induced by the classical adiabatic force,
${\bf f}^0 = \dot{\bf p}^\mathrm{0,cl}$.

%
\section{Conclusion}
\setcounter{equation}{0} \setcounter{subsubsection}{0}
%

\subsubsection{Limitations of the Present Results}

Several approximations and simplifications limit
the quantitative reliability of the  present simulations.
First is the reliance upon the Lennard-Jones pair potential,
which is exact for the long range attraction,
but which approximates the functional form and strength
of the short range repulsion.
Second, is the neglect of many body interactions,
of which the leading order Axilrod-Teller three-body potential
is predominantly repulsive at short-range.
Third is the neglect of the commutation function,
which is short-ranged and repulsive.
The commutation function
ultimately reflects the Heisenberg uncertainty principle
and the zero point energy,
and it means that at low temperatures $^4$He remains a liquid
where otherwise it would be expected to be a solid.
Whereas Bose-Einstein condensation is a non-local phenomenon
that is insensitive to short-range effects,
the viscosity itself depends on short-range interactions.
In consequence the present simulations
are approximate in that they neglect these three effects;
the change in viscosity in going from the classical to the quantum liquid
is more reliable than the individual values.
A related point is that the saturation liquid density
used in the present QSMD simulations
is the value for Lennard-Jones $^4$He
obtained from classical Monte Carlo simulations (Attard 2023 table~8.3).

The fourth limitation of the present simulations
is that they invoke only pure momentum loops.
Although exact for ideal bosons.
for the more realistic case of interacting bosons,
mixed permutation loops,
where the bosons can be in different momentum states,
contribute significantly
on the high temperature side of the $\lambda$-transition.
Pure permutation loops can be expected to dominate
on the low temperature side of the $\lambda$-transition
(Attard 2023 section~9.3).
The $\lambda$-transition and Bose-Einstein condensation
for ideal bosons occurs for $\rho_\mathrm{c,id} \Lambda_\mathrm{c,id}^3 = 2.612$
(F. London 1938, Pathria 1972, Attard 2022, 2023).
The present results
for saturated liquid  Lennard-Jones $^4$He
span the ideal boson transition
and end at $\rho \Lambda^3 = 2.70$ at $T^*=0.55$.
The present treatment of wave function symmetrization is that of ideal bosons,
and the thermodynamic state point of that treatment
includes the regime of Bose-Einstein  condensation for ideal bosons.

A fifth limitation is the calculation of the viscosity
by integrating the time correlation function
over the  continuous occupancy trajectory.
Instead
one should weight each time step on the trajectory by the ratio
of the exact to the umbrella transition weight.
Unlike the umbrella sampling of the phase space points themselves
where the exact probability can be calculated
from the exact occupancy and permutation entropy,
the exact transition probability requires the gradient of the permutation
entropy, which is problematic to obtain (appendix~\ref{Sec:ConstEntTraj}).

The continuous occupancy and umbrella sampling method
poses statistical limitations.
The quantum liquid is more challenging
than the classical liquid, and the statistical error
is much larger for a given total time.
Also the length of the  time step needs to be smaller
to solve  the decoherent quantum trajectory accurately.
It appears that the numerical procedures are quite sensitive
to the functional form and to the value of the parameters
used for the continuous occupancy.
Although the average values are not affected
by the chosen form and parameter values
for the continuous occupancy,
the length of the simulation required to produce reliable results is.

\subsubsection{New Concepts}

Previous work has shown that entanglement with the environment or reservoir
decoheres the subsystem wave function
and in a formally exact sense enables the quantum state of an open system
to be characterized by simultaneous values
of the particles' positions and momenta,
summing over which gives statistical averages (Attard 2018b, 2021, 2023).
A similar decoherence has been obtained with the environmental selection approach
that has been applied to the quantum measurement problem
(Zeh 2001, Zurek 2003, Schlosshauer, 2005).
The main differences are that
the phase space analysis is in the context
of quantum statistical mechanics
and it provides an actual probability distribution.
The classical phase space formulation is a form of quantum realism.

The major conceptual contribution of the present paper
is the decoherent quantum equations of motion,
$\dot{\bf \Gamma}^0 = - T \nabla^\dag S({\bf \Gamma})$.
This reduces to Hamilton's equations in the classical limit,
and makes a phase space average equals the simple time average
over the trajectory.
The decoherent quantum equations of motion
ensure the constancy of the exact equilibrium probability density
of the open quantum system.
The added stochastic dissipative thermostat confers stability to the evolution
(appendix~\ref{Sec:Stab}).

The existence of molecular trajectories (section~\ref{Sec:Traj})
is a challenge to conventional quantum mechanics.
The Copenhagen interpretation
says that in a closed system the position and momentum of a particle
do not exist between measurements,
and that there is no such thing as a particle trajectory.
The state of the closed subsystem is believed
to come into existence only when it is observed.
The Copenhagen interpretation is an anti-realist theory.

The present analysis and calculations are for an open quantum subsystem.
The exchange of energy with the heat reservoir
entangles the wave function and causes it to collapse.
This allows the state of the subsystem to be specified
by a point in classical phase space
(Attard  2018b, 2021, 2023).
The present trajectories are a result of the equations of motion
that correspond to a transition probability
that preserves the equilibrium probability distribution
during the evolution of the subsystem.
This is a fundamental requirement of probability theory (Attard 2012).
The specification of the state of the subsystem
as a point in classical phase space,
and the existence of a trajectory,
means that this treatment of an open quantum system
may be classified as a realist theory:
The state of the subsystem exists whether or not an observer
makes a measurement.

A consequence of the present decoherent quantum equations of motion
is that the velocity does not equal
the momentum divided by the  mass
when changes in momentum state occupancy are involved.
Changes in occupancy
can be precluded by jumps in position.
In the present calculations these jumps are exceedingly small
compared to the size of $^4$He.
The jumps are likely to be larger and more abrupt
in the discrete occupancy limit, $\kappa \rightarrow \infty$
(cf.\ appendix~\ref{Sec:DiscOccTraj}).
The spikes containing velocity reversal
cancel in the classical limit,
but  in the quantum regime step jumps certainly will not.
Perhaps one might regard the jumps as a remnant
of the underlying quantum nature of a closed subsystem,
which, according to the Copenhagen interpretation,
precludes a continuous trajectory.
Although the decoherent quantum equations of motion are deterministic,
the jumps would appear to be random in comparison
to the smooth classical trajectory.

In the present results
the classical limit emerges naturally
from the decoherent quantum equations of motion.
In the classical extreme the momentum state occupancy is only ever 0 or 1,
the velocity reversal spikes predominate,
and, since the two components of such jumps have opposite sign,
in the discrete occupancy limit, $\kappa \rightarrow \infty$,
these likely become superimposed and cancel
so that they do not actually effect the trajectory.
This was found analytically to be the case
for discrete occupancy (appendix~\ref{Sec:DiscOccTraj}).
That is, for uncondensed bosons
the change in a momentum state occurs without even a speed bump.
Hence in the classical limit
the boson decoherent quantum trajectory becomes continuous,
and Hamilton's adiabatic classical equations of motion emerge.


\subsubsection{Algorithmic Advances}

Two practical problems had to be solved
in order to implement the present decoherent quantum equations of motion
on a computer,
both for the present study of $^4$He,
and for the simulation of open quantum systems more generally.
The first was
to combine the continuum momenta of classical phase space
with the discrete momentum eigenvalues of the quantum system.
The second was to provide a viable procedure
that gives the gradient of the permutation entropy.
The discrete momentum eigenvalues
are necessary to obtain the momentum state occupancy
and hence the permutation entropy for quantum statistics.
The second problem was solved by defining a continuous form of occupancy
with a computable gradient,
which was  treated as a form of umbrella sampling over the trajectory
from which the exact phase space averages could be obtained.

Combining the momentum continuum with the quantized eigenvalues
gives the transitions between momentum states.
With this and the decoherent quantum equations of motion
one has the generalization of Schr\"odinger's equation
to an open quantum system.

The quantum stochastic molecular dynamics (QSMD) simulation method
enables the description of macroscopic quantum condensed matter systems
at the molecular level.
The continuous occupancy
presents both a challenge and an opportunity for further developments.

The decoherent quantum equations of motion  formally apply to fermions
with the permutation entropy replaced
by the Fermi exclusion principle.
An efficient form of umbrella sampling
lies in the details.

\subsubsection{Results for Liquid Helium}

Beyond the quantum statistical theory
and computational algorithms developed here
lie the specific results obtained for liquid helium
at temperatures encompassing the $\lambda$-transition.
Perhaps the most significant of these are the results for the viscosity.

As far as I am aware these are the first realistic results
for the superfluidity of interacting bosons.
I do not count the phenomenological two-fluid theory of Tisza (1938),
which uses F. London's (1938) analysis of Bose-Einstein condensation
based on ideal (non-interacting) boson statistics.
Nor do I count Landau's (1941) phonon-roton phenomenological theory,
which is based solely on the first excited energy state.
This might perhaps be relevant at 1~fK,
but not at the temperatures of the $\lambda$-transition
where superfluidity has actually been measured.
Landau never accepted Bose-Einstein condensation
as the origin of the $\lambda$-transition or of superfluidity,
and so it is a little hard to understand
how anyone can simultaneously hold to be true
both Bose-Einstein condensation
and Landau's theory of superfluidity.

The present results show that the reduction
in the viscosity from its classical value
coincides with the condensation of bosons
into multiple multiply-occupied low-lying momentum states.
Defining
uncondensed bosons to be those in few-occupied momentum states,
and condensed bosons to be those in highly occupied momentum states,
one could perhaps argue for a form of Tisza's (1938)
two-fluid theory,
namely a linear combination of zero viscosity of the condensed bosons
and the non-zero classical value of the uncondensed bosons.
However, like Feynman (1954) I see this as unrealistic
since there is no sharp delineation between the two.
The present results for a spectrum of momentum state occupancies
 underscore the difficulty in turning Tisza's (1938) two-fluid theory
into something with quantitative predictive value.

One thing that is certain is that the ground momentum state
does not play the dominant role in superfluidity
that  has hitherto been assumed.
The present results show a reduction in viscosity
at temperatures close to the $\lambda$-transition temperature
even though the ground momentum state is not the sole highly occupied state,
or even at every instant the most occupied state.
Furthermore the occupancy of the ground momentum state
does not become macroscopic in the thermodynamic limit.
Since superfluidity can be observed and measured in the laboratory
it must be a macroscopic phenomenon.
The number of condensed bosons by the present definition
becomes macroscopic in the thermodynamic limit,
whereas the number in the ground momentum state does not.

\subsubsection{Molecular Interpretation of Superfluidity}
\label{Sec:Landau}

And now to Landau's objection
to Bose-Einstein condensation as the origin of superfluidity:
`nothing would prevent atoms in a normal state from colliding with
excited atoms, ie.\ when moving through the liquid they would
experience a friction and there would be no superfluidity at all' (Landau 1941).

The fundamental conceptual conclusion
that can be drawn from the present equations of motion
is that the total entropy is conserved on the decoherent quantum trajectory.
This means that any change in permutation entropy
in a collision must be compensated by an equal and opposite change
in the reservoir entropy.
(The decoherent quantum equations do \emph{not} maximize the entropy.)
If an upcoming change in permutation entropy,
on, say, a classical trajectory,
would be so large that it could not be so compensated,
than the trajectory itself must shift to avoid the collision.

\begin{figure}[t]
\centerline{ \resizebox{8cm}{!}{ \includegraphics*{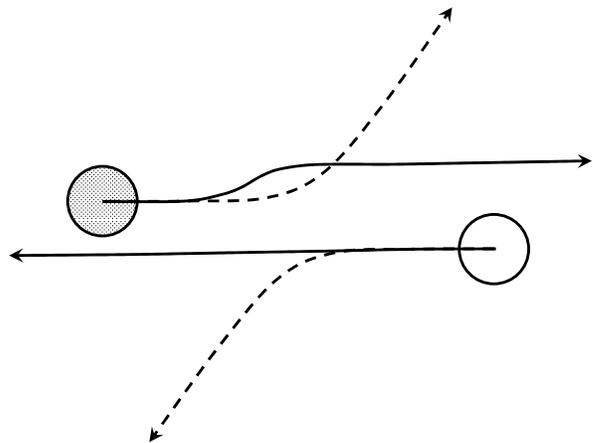} } }
\caption{\label{Fig:coll}
Schematic based on the trajectory in figure~\ref{Fig:traj4}
of a collision in the $xy$-plane
between a condensed boson (filled)
and an uncondensed boson (empty)
in the classical case (dashed curves, ballistic collision)
and in the quantum case (solid curves, side-step collision).
}
\end{figure}

The quantum term in the decoherent quantum equations of motion (\ref{Eq:dotG0}),
$\dot{\bf q}^\mathrm{0,qu} = -T \nabla_p S^\mathrm{perm}({\bf p})$,
has the effect of eliminating the shear viscosity in the condensed regime,
as the following argument shows.
Consider the glancing collision between two repulsive particles
(figure~\ref{Fig:coll}).
The sketch is based on the trajectory computed
in figure~\ref{Fig:traj4}
where the uncondensed boson
is little affected by the collision (not shown).

Suppose that the collision occurs in superfluid flow from left to right,
with the highly occupied momentum states being
highly aligned and clustered about some non-zero magnitude.
On approach of the pair close enough to feel the repulsive part
of the interaction potential,
the classical adiabatic part of the collision attempts to rotate
the momentum of the condensed boson.
But this would knock it into a few-occupied state,
thereby reducing the permutation entropy.

Instead the decoherent quantum term
increases the collision offset (ie.\ impact parameter)
such that the distance of closest approach
is larger than would occur on the classical trajectory.
Prior to the lateral jump
the lateral component of the repulsive force
causes the boson to approach the lateral momentum boundary,
the crossing of which would change the permutation entropy.
The jump to a region of zero force precludes
the change in momentum state and conserves the entropy.

This description of zero change in permutation entropy
while the condensed particle follows a trajectory of zero force
is a simplified picture that likely holds when the change in momentum state
would lead to a large change in permutation entropy.
This is confirmed by many of the jumps
shown in figures~\ref{Fig:traj}--\ref{Fig:traj4},
which do not result in a change in momentum state
because the jump reverses the sign of the force.

The decoherent quantum equations of motion show that
in an open quantum system the change in position
can occur independently of the momentum.
The decoherent quantum term can do this because
it is not bound by the classical notion that the rate of change of position
must equal the momentum divided by the particle mass.
In other words, the velocity of the quantum trajectory in position space
is not equal to the momentum  divided by mass
when changes in permutation entropy are involved.

After the collision,
which is time-reversible,
the condensed boson remains within
the envelope of highly occupied momentum states for the superfluid flow.
The fact that the distance of closest approach
is larger in the quantum case than in the classical case means
that the transfers of momenta longitudinally and laterally
are likely zero.
In figures~\ref{Fig:traj4} and \ref{Fig:coll},
the lateral momentum state of the condensed boson
does not change in the collision,
and since total momentum is conserved on a decoherent quantum trajectory,
neither does that of the uncondensed boson.
Since the shear viscosity is a direct measure of longitudinal momentum
transferred laterally,
it must be zero in the quantum condensed regime.

This answers Landau's objection.
It explains at the molecular level how a collision
between a condensed and an uncondensed boson
does not result in the lateral transfer of longitudinal momentum,
and why the shear viscosity vanishes.
It shows, qualitatively,
why Bose-Einstein condensation gives rise to superfluidity.
The quantitative molecular measure of the effect
is provided by the viscosity given
by the present quantum stochastic molecular dynamics algorithm.

\section*{References}


\begin{list}{}{\itemindent=-0.5cm \parsep=.5mm \itemsep=.5mm}

\item
Abramowitz M and Stegun IA 1972
\emph{Handbook of Mathematical Functions}
(New York: Dover)

\item 
Attard  P 2002
\emph{Thermodynamics and statistical mechanics:
equilibrium by entropy maximisation}
(London: Academic Press)

\item 
Attard  P 2012
\emph{Non-equilibrium thermodynamics and statistical mechanics:
Foundations and applications}
(Oxford: Oxford University Press)

\item
Attard P 2018a
Shear thinning in Lennard-Jones fluids by stochastic dissipative
molecular dynamics simulation
arXiv:1810.12523

\item 
Attard P 2018b
Quantum statistical mechanics in classical phase space. expressions for
the multi-particle density, the average energy, and the virial pressure
arXiv:1811.00730

\item 
Attard P  2021
\emph{Quantum Statistical Mechanics in Classical Phase Space}
(Bristol: IOP Publishing)

\item 
Attard P 2022
Bose-Einstein condensation, the lambda transition, and superfluidity for
interacting bosons
arXiv:2201.07382

\item 
Attard  P 2023
\emph{Entropy beyond the second law.
Thermodynamics and statistical mechanics
for equilibrium, non-equilibrium, classical, and quantum systems}
(Bristol: IOP Publishing, 2nd edition)

\item 
Balibar S (2014)
Superfluidity: How Quantum Mechanics Became Visible.
In:
Gavroglu, K. (eds)
\emph{History of Artificial Cold, Scientific,
Technological and Cultural Issues.}
Boston Studies in the Philosophy
and History of Science, {\bf 299}
(Dordrecht: Springer)

\item 
Balibar S 2017
Laszlo Tisza and the two-fluid model of superfluidity
\emph{C.\ R.\ Physique} {\bf 18}  586

\item 
Feynman R P 1954
Atomic theory of the two-fluid model of liquid helium
\emph{Phys.\ Rev.}\ {\bf 94} 262

\item
Green MS 1954
Markoff random processes and the statistical mechanics of time-dependent phenomena.
II. Irreversible processes in fluids.
\emph{J.\ Chem.\ Phys.}\ {\bf 23}, 298

\item
Kubo R 1966
The fluctuation-dissipation theorem
\emph{Rep.\ Prog.\ Phys.}\ {\bf 29} 255

\item  
London F 1938
The $\lambda$-phenomenon of liquid helium and the Bose-Einstein degeneracy
\emph{Nature} {\bf 141} 643

\item 
Merzbacher E 1970
\emph{Quantum Mechanics} 2nd edn
(New York: Wiley)

\item 
Messiah A 1961 \emph{Quantum Mechanics}
(Vol 1 and 2) (Amsterdam: North-Holland)

\item
Onsager L (1931)
Reciprocal Relations in Irreversible Processes. I.
\emph{Phys.\ Rev.}\ {\bf 37} 405.
Reciprocal Relations in Irreversible Processes. II.
\emph{Phys.\ Rev.}\ {\bf 38} 2265

\item
Pathria R K 1972
\emph{Statistical Mechanics} (Oxford: Pergamon Press)

\item  
Tisza L 1938
Transport phenomena in helium II
\emph{Nature} {\bf 141} 913

\item 
Landau L D 1941
Theory of the superfluidity of helium II
\emph{Phys.\ Rev.}\ {\bf 60} 356

\item  
Schlosshauer M 2005
Decoherence, the measurement problem, and interpretations of quantum
mechanics
arXiv:quant-ph/0312059v4

\item 
van Sciver  S W 2012
\emph{Helium Cryogenics}
2nd edn (New York: Springer)

\item  
Zeh HD 2001
\emph{The Physical Basis of the Direction of Time} 4th edn
(Berlin: Springer)

\item  
Zurek WH 2003
Decoherence, einselection,
and the quantum origins of the classical
arXiv:quant-ph/0105127v3

\end{list}



\appendix
%
\section{Stability of the Equilibrium Probability Density}
\label{Sec:Stab}
\setcounter{equation}{0} \setcounter{subsubsection}{0}
\renewcommand{\theequation}{\Alph{section}.\arabic{equation}}
%

This appendix shows that the equilibrium probability density
is stable to small perturbations during its evolution
according to the stochastic dissipative decoherent quantum
equations of motion.

The cross-entropy measures the distance between
an approximate and the exact probability distribution.
It is defined as (Attard 2002 equation~(10.22))
\begin{eqnarray}
S[\tilde \wp,\wp]
& = &
\int \mathrm{d} {\bf \Gamma}\;
\tilde \wp({\bf \Gamma})  \ln \frac{\tilde\wp({\bf \Gamma}) }{\wp({\bf \Gamma}) }
 \\ & = & \nonumber
\int \mathrm{d} {\bf \Gamma}\;  \wp({\bf \Gamma})
\left\{
\frac{\tilde\wp({\bf \Gamma}) }{\wp({\bf \Gamma}) }
\ln \frac{\tilde\wp({\bf \Gamma}) }{\wp({\bf \Gamma}) }
-
\frac{\tilde\wp({\bf \Gamma}) }{\wp({\bf \Gamma}) }
+ 1 \right\}.
\end{eqnarray}
The second equality follows because the probabilities are normalized to unity.
Since $x \ln x \ge x - 1$, the integrand is non-negative,
which means that the cross-entropy vanishes if and only if
$\tilde \wp = \wp$.

With the approximate probability density time-dependent,
the rate of change of the distance is
\begin{eqnarray}
\frac{\partial  S[\tilde \wp,\wp]}{\partial t}
& = &
\int \mathrm{d} {\bf \Gamma}\;
\frac{\partial \tilde \wp({\bf \Gamma},t) }{\partial t}
\ln \frac{\tilde\wp({\bf \Gamma},t) }{\wp({\bf \Gamma}) } ,
\end{eqnarray}
This follows because normalization implies
that $\int \mathrm{d} {\bf \Gamma}\;
{\partial \tilde \wp({\bf \Gamma},t) }/{\partial t} = 0$.

The  stochastic dissipative decoherent quantum equations of motion
given in the text, equation~(\ref{Eq:SDdQ-EoM}),
have the same functional form as those derived in the classical case
from the second entropy formulation  of non-equilibrium thermodynamics
(Attard 2012).
Hence the rate of change of the approximate probability density
due to these equations of motion
is given by
(Attard 2012 equation~(7.102))
\begin{eqnarray}
\frac{\partial  \tilde \wp({\bf \Gamma},t) }{\partial t}
& = &
\frac{-1}{\Delta_t}
[ \nabla_p \cdot \overline {\bf R}_p ] \tilde \wp({\bf \Gamma},t)
+ \frac{\sigma^2}{2\Delta_t} \nabla_p^2 \tilde \wp({\bf \Gamma},t)
\nonumber \\ && \mbox{ }
- \left[ \dot{\bf \Gamma}^0
+ \frac{1}{\Delta_t} \overline {\bf R}_p  \right]
\cdot \nabla \tilde \wp({\bf \Gamma},t) .
\end{eqnarray}
This is a form of the Fokker-Planck equation.
In the present paper the dissipative force is
$\overline {\bf R}_p  = (\sigma^2/2k_\mathrm{B}) \nabla_p S({\bf \Gamma})$,
the variance squared is
$\langle \tilde {\bf R}_p \tilde {\bf R}_p \rangle = \sigma^2\mathrm{I}_{pp}$,
and $\dot{\bf \Gamma}^0 = - T \nabla^\dag S$.

Write the approximate probability as
$\tilde\wp({\bf \Gamma},t)
= \tilde Z(t)^{-1} \exp \tilde S({\bf \Gamma},t)/k_\mathrm{B}$,
with
$ \tilde S({\bf \Gamma},t) - S({\bf \Gamma})
= k_\mathrm{B} \varepsilon({\bf \Gamma},t)$.
To linear order in $\varepsilon$ the partial time derivative is
\begin{eqnarray}
\lefteqn{
\frac{\partial  \tilde \wp({\bf \Gamma},t) }{\partial t}
}  \\
& = &
\frac{-1}{\Delta_t}
\frac{\sigma^2}{2k_\mathrm{B}} [  \nabla_p^2   S] \tilde \wp({\bf \Gamma},t)
+ T \nabla^\dag S \cdot [ \nabla S/k_\mathrm{B} + \nabla \varepsilon ]
\tilde \wp({\bf \Gamma},t)
\nonumber \\ &  & \mbox{ }
- \frac{1}{\Delta_t} \frac{\sigma^2}{2k_\mathrm{B}}  \nabla_p  S
\cdot [ \nabla_p S/k_\mathrm{B} + \nabla_p \varepsilon ]
\tilde \wp({\bf \Gamma},t)
\nonumber \\ &  & \mbox{ }
+ \frac{\sigma^2}{2\Delta_t}
\Big[ \nabla_p^2 S/k_\mathrm{B} + \nabla_p^2 \varepsilon
+ \big(\nabla_p S/k_\mathrm{B}
+ \nabla_p \varepsilon \big)^2  \Big]
 \tilde \wp({\bf \Gamma},t)
\nonumber \\ & = &
T (\nabla^\dag S) \cdot (\nabla \varepsilon )\tilde \wp({\bf \Gamma},t)
- \frac{1}{\Delta_t} \frac{\sigma^2}{2k_\mathrm{B}}  (\nabla_p  S)
\cdot  ( \nabla_p \varepsilon) \tilde \wp({\bf \Gamma},t)
\nonumber \\ &  & \mbox{ }
+ \frac{\sigma^2}{2\Delta_t}
\Big[  \nabla_p^2 \varepsilon
+ \frac{2}{k_\mathrm{B}} \nabla_p S\cdot\nabla_p \varepsilon
+ \nabla_p \varepsilon\cdot \nabla_p \varepsilon  \Big]
 \tilde \wp({\bf \Gamma},t)
\nonumber \\ & = &
\left\{
T \nabla^\dag S \cdot \nabla \varepsilon
+ \frac{1}{\Delta_t} \frac{\sigma^2}{2k_\mathrm{B}}
\nabla_p  S \cdot \nabla_p \varepsilon
+ \frac{\sigma^2}{2\Delta_t} \nabla_p^2 \varepsilon
\right\} \wp({\bf \Gamma}) .
\nonumber
\end{eqnarray}
Hence
\begin{eqnarray}
\frac{\partial  S[\tilde \wp,\wp]}{\partial t}
& = &
\int \mathrm{d} {\bf \Gamma}\;
 \wp({\bf \Gamma})
\left[ \varepsilon + \frac{Z}{\tilde Z(t)} \right]
\left\{\rule{0cm}{.5cm} 
T \nabla^\dag S \cdot \nabla \varepsilon
\right. \nonumber \\ &  & \left. 
+ \frac{\sigma^2}{2k_\mathrm{B}\Delta_t}
\nabla_p  S \! \cdot \! \nabla_p \varepsilon
+ \frac{\sigma^2}{2\Delta_t} \nabla_p^2 \varepsilon
\right\} .
\end{eqnarray}
Integrating the term proportional to the temperature by parts gives
\begin{eqnarray}
\lefteqn{
\int \mathrm{d} {\bf \Gamma}\;
\wp({\bf \Gamma})
\left[ \varepsilon + \frac{Z}{\tilde Z(t)} \right]
T (\nabla^\dag S) \cdot (\nabla \varepsilon )
} \nonumber \\
& = &
\frac{k_\mathrm{B} T}{2}
\int \mathrm{d} {\bf \Gamma}\;
\big(\nabla^\dag  \wp({\bf \Gamma})\big)
 \cdot \nabla\left[ \varepsilon + \frac{Z}{\tilde Z(t)} \right]^2
 \nonumber \\ & = &
\frac{-k_\mathrm{B} T}{2}
\int \mathrm{d} {\bf \Gamma}\;
 \wp({\bf \Gamma})
\nabla^\dag \cdot \nabla\left[ \varepsilon + \frac{Z}{\tilde Z(t)} \right]^2
 \nonumber \\ & = &
 0.
\end{eqnarray}
Similarly the dissipative term yields
\begin{eqnarray}
\lefteqn{
\int \mathrm{d} {\bf \Gamma}\;
\wp({\bf \Gamma})
\left[ \varepsilon + \frac{Z}{\tilde Z(t)} \right]
 \frac{\sigma^2}{2k_\mathrm{B}\Delta_t}
 (\nabla_p  S) \cdot  ( \nabla_p \varepsilon)
}  \\
& = &
 \frac{\sigma^2}{4\Delta_t}
\int \mathrm{d} {\bf \Gamma}\;
\big(\nabla_p  \wp({\bf \Gamma})\big)
 \cdot \nabla_p \left[ \varepsilon + \frac{Z}{\tilde Z(t)} \right]^2
 \nonumber \\ & = &
 \frac{-\sigma^2}{4\Delta_t}
\int \mathrm{d} {\bf \Gamma}\;
 \wp({\bf \Gamma})
\nabla_p^2 \left[ \varepsilon + \frac{Z}{\tilde Z(t)} \right]^2
 \nonumber \\ & = &
 \frac{-\sigma^2}{2\Delta_t}
\int \mathrm{d} {\bf \Gamma}\;
 \wp({\bf \Gamma})
 \nabla_p \cdot\left\{ \left[ \varepsilon + \frac{Z}{\tilde Z(t)} \right]
    \nabla_p    \varepsilon \right\}
  \nonumber \\ & = &
 \frac{-\sigma^2}{2\Delta_t}
\int \mathrm{d} {\bf \Gamma}\;
 \wp({\bf \Gamma})
 \left\{ \left[ \varepsilon + \frac{Z}{\tilde Z(t)} \right]
 \nabla_p^2 \varepsilon
 + \nabla_p \varepsilon  \cdot \nabla_p \varepsilon   \right\}.
 \nonumber
\end{eqnarray}
The first term here cancels with the final term above and one is left with
\begin{equation}
\frac{\partial  S[\tilde \wp,\wp]}{\partial t}
=
 \frac{-\sigma^2}{2\Delta_t}
\int \mathrm{d} {\bf \Gamma}\;
 \wp({\bf \Gamma}) \,
 ( \nabla_p \varepsilon ) \cdot ( \nabla_p \varepsilon ) .
\end{equation}
Going forward in time, $\Delta_t > 0$, this is negative,
which is to say that the distance
from the exact probability distribution decreases.
Hence
the equilibrium quantum probability density in classical phase space
is stable to small perturbations
when it evolves according to
the stochastic dissipative decoherent quantum
equations of motion.

%
\section{Second Order Equations of Motion} \label{Sec:2ndOEoM}
\setcounter{equation}{0} \setcounter{subsubsection}{0}
\renewcommand{\theequation}{\Alph{section}.\arabic{equation}}
%

The second order decoherent quantum equations of motion are
\begin{eqnarray}
\lefteqn{
{\bf \Gamma}^0(t+\tau|{\bf \Gamma}(t),t)
}  \\
& = &
{\bf \Gamma}(t) + \tau \dot{\bf \Gamma}^0(t)
+ \frac{\tau^2}{2} \dot{\bf \Gamma}^0(t) \cdot \nabla \dot{\bf \Gamma}^0(t)
\nonumber \\ & = &
{\bf \Gamma}(t) - \tau T \nabla^\dag S(t)
+ \frac{\tau^2T^2 }{2} \nabla^\dag S(t) \cdot \nabla \nabla^\dag S(t) ,\nonumber
\end{eqnarray}
or
\begin{eqnarray}
\lefteqn{
{\bf q}^0(t+\tau|{\bf \Gamma}(t),t)
}  \nonumber \\
& = &
{\bf q}(t)
+ \frac{\tau}{m} {\bf p}(t)
 - \tau T \nabla_p  S^\mathrm{perm}({\bf p}(t))
 \nonumber \\ && \mbox{ }
+ \frac{\tau^2}{2} {\bf f}^0(t) \cdot \nabla_p\!
\left[ \frac{1}{m}{\bf p}(t) -  T \nabla_p  S^\mathrm{perm}({\bf p}(t)) \right]\!,
\end{eqnarray}
and
\begin{eqnarray}
\lefteqn{
{\bf p}^0(t+\tau|{\bf \Gamma}(t),t)
}  \nonumber \\
& = &
{\bf p}(t)
+  \tau  {\bf f}^0(t)
 \nonumber \\ && \mbox{ }
+ \frac{\tau^2}{2}\!
\left[ \frac{1}{m}{\bf p}(t) -  T \nabla_p  S^\mathrm{perm}({\bf p}(t)) \right]\!
\cdot \nabla_q {\bf f}^0(t) .
\end{eqnarray}

To these the thermostat terms are simply added.
These are first order, and their strength  $\sigma$
can be chosen independent of the length of the time step
and of the order of the decoherent quantum equations of motion.


The elements of the dyadic,
$\nabla_{p,j} \nabla_{p,k} S^\mathrm{perm}({\bf p})$,
are required.
Since the permutation entropy is the sum over individual single particle states,
the second derivative is non-zero only for states
that are affected by both bosons simultaneously,
which is to say ${\bf a}_j'' = {\bf a}_k''$.
Here ${\bf a}_j'' = {\bf a}_j
+ \sum_\alpha r_{j\alpha}  s_{j\alpha} \widehat{\bm \alpha}$
is one of the eight states that have weight from momentum boson
${\bf p}_j \in {\bf a}_j $,
and similarly for
 ${\bf a}_k'' = {\bf a}_k
+ \sum_\gamma r_{k\gamma}  s_{k\gamma} \widehat{\bm \gamma}$.
Hence
\begin{eqnarray}
\lefteqn{
\nabla_{p,j} \nabla_{p,k} S^\mathrm{perm}({\bf p})
} \nonumber \\
& = &
k_\mathrm{B}
\sum_{ {\bf r}_j, {\bf r}_{k} }
\delta_{{\bf a}_j'',{\bf a}_k''} \left\{
\frac{\partial \ln \Gamma( {\cal N}_{{\bf a}_j''} + 1 )
}{ \partial \eta_{{\bf a}_j''} }
\;
[\nabla_{p,j} \nabla_{p,j}\Phi_{j}^{({\bf r}_{j})}] \delta_{jk}
\right. \nonumber \\ && \mbox{ } \left.
+
\frac{\partial^2 \ln \Gamma( {\cal N}_{{\bf a}_j''} + 1 )
}{ \partial \eta_{{\bf a}_j''}^2 }
\;
[\nabla_{p,j}\Phi_{j}^{({\bf r}_{j})}]\,
[\nabla_{p,k} \Phi_{k}^{({\bf r}_{k})}]
\right\}.
\end{eqnarray}
Recall that
$ \Phi_{j}^{({\bf r}_{j})} \equiv
\Phi_{jx}^{(r_{jx})} \Phi_{jy}^{(r_{jy})} \Phi_{jz}^{(r_{jz})} $
with
$\Phi_{j\alpha}^{(r_{j\alpha})}
\equiv
\overline r_{j\alpha} \phi_{j\alpha} + r_{j\alpha} \overline \phi_{j\alpha} $.
One has
\begin{eqnarray}
\lefteqn{
\nabla_{p,j\alpha} \nabla_{p,j\gamma}\Phi_{j}^{({\bf r}_{j})}
}  \\
& = &
\nabla_{p,j\alpha} \nabla_{p,j\gamma}
[\overline r_{jx} \phi_{jx} + r_{jx} \overline \phi_{jx}]
[\overline r_{jy} \phi_{jy} + r_{jy} \overline \phi_{jy}]
\nonumber \\ && \mbox{ } \times
[\overline r_{jz} \phi_{jz} + r_{jz} \overline \phi_{jz}]
\nonumber \\ & = &
\nabla_{p,j\alpha} [ 1- 2 r_{j\gamma}]
\phi_{j\gamma}'
\Phi_{j\gamma'}^{(r_{j\gamma'})} \Phi_{j\gamma''}^{(r_{j\gamma''})}
, \;\; \mbox{all $\gamma$ different}
\nonumber \\ & = &
\left\{ \begin{array}{ll}
[ 1- 2 r_{j\gamma}] \phi_{j\gamma}''
\Phi_{j\gamma'}^{(r_{j\gamma'})} \Phi_{j\gamma''}^{(r_{j\gamma''})},
& \alpha = \gamma, 
\\
\,\! [ 1- 2 r_{j\alpha} ] [ 1- 2 r_{j\gamma}]    
 \phi_{j\alpha}' \phi_{j\gamma}' \Phi_{j\nu}^{(r_{j\nu})}
& \alpha \ne \gamma \ne \nu ,
\end{array} \right.\nonumber
\end{eqnarray}
where the $\gamma$ are all different.
Also
\begin{eqnarray}
\lefteqn{
[\nabla_{p,j\alpha} \Phi_{j}^{({\bf r}_{j})}]\,
[\nabla_{p,k\gamma} \Phi_{k}^{({\bf r}_{k})}]
} \nonumber \\
& = &
[ 1- 2 r_{j\alpha}]
\phi_{j\alpha}' \Phi_{j\alpha'}^{(r_{j\alpha'})} \Phi_{j\alpha''}^{(r_{j\alpha''})}
\nonumber \\ &  & \mbox{ } \times
[ 1- 2 r_{k\gamma}]
\phi_{k\gamma}' \Phi_{k\gamma'}^{(r_{k\gamma'})}
\Phi_{k\gamma''}^{(r_{k\gamma''})} .
\end{eqnarray}
A neighbor table in momentum space
facilitates the computation of the dyadic terms.

%
\section{Discrete Occupancy Trajectory} \label{Sec:DiscOccTraj}
\setcounter{equation}{0} \setcounter{subsubsection}{0}
\renewcommand{\theequation}{\Alph{section}.\arabic{equation}}
%

This appendix analyzes the decoherent equations of motion
to first order over a time step
for the exact case that the occupancy is discrete.
It is of some interest to see how the formal expression
yields the classical limit.

For momentum configuration
${\bf p} =\{{\bf p}_1,{\bf p}_2, \ldots ,{\bf p}_N\}$,
the occupancy of the momentum state ${\bf a} = {\bf n} \Delta_p$
is
\begin{eqnarray}
N_{\bf a}({\bf p})
& = &
\sum_{j=1}^N \prod_{\alpha=x,y,z}
\Theta\big( p_{j\alpha} - ( a_{\alpha} - \Delta_p/2) \big) \,
\nonumber \\ && \mbox{ } \times
\Theta\big(  a_{\alpha} + \Delta_p/2 - p_{j\alpha} \big) ,
\end{eqnarray}
where $\Theta(x)$ is the Heaviside unit step function.
The gradient of this is
\begin{eqnarray}
\lefteqn{
\nabla_{p,j\alpha} N_{\bf a}({\bf p})
}  \\
& = &
\Big\{
\delta\big( p_{j\alpha}- (a_\alpha-\Delta_p/2)\big)
-
\delta\big( p_{j\alpha}- (a_\alpha+\Delta_p/2)\big)
\Big\}
\nonumber \\ && \mbox{ } \times
\prod_{\gamma\ne \alpha}
\Theta\big( p_{j\gamma} - ( a_{\gamma} - \Delta_p/2) \big) \,
\Theta\big(  a_{\gamma} + \Delta_p/2 - p_{j\gamma} \big) .
 \nonumber
\end{eqnarray}

The permutation entropy is
$ S^\mathrm{perm}({\bf p})
= k_\mathrm{B}  \sum_{\bf a} \ln N_{\bf a}({\bf p}) !$,
and its gradient has components
\begin{eqnarray}
\nabla_{p,j\alpha} S^\mathrm{perm}({\bf p})
& = &
s_{j\alpha} k_\mathrm{B}
\ln \left[ \frac{ N_{{\bf a}'}({\bf p}) +1 }{N_{{\bf a}}({\bf p}) } \right]
\nonumber \\ && \mbox{ } \times
\delta\big( p_{j\alpha}- (a_\alpha+a'_{\alpha})/2\big) .
\end{eqnarray}
Here ${\bf p}_j(t) \in {\bf a}$,
${\bf p}_j^{0,\mathrm{cl}}(t+\tau) \in {\bf a}'$,
and $s_{j\alpha} \equiv ( a_\alpha' - a_\alpha)/\Delta_p = 0 \mbox{ or } {\pm 1} $.
The transition state ${\bf a}'$  is one of the 26 neighboring states
to  ${\bf a}$.
This holds for each direction individually,
with the gradient being zero in directions
where there is no change in momentum state.
This assumes that transitions are a rare event
that can be treated independently.

The decoherent quantum evolution of the position
to linear order in the time step is
\begin{eqnarray} \label{Eq:dot-q_ja^0qu}
\lefteqn{
{q}_{j\alpha}^{0}(t+\tau|{\bf \Gamma},t)
}  \\
& = &
{q}_{j\alpha}^{0,\mathrm{cl}}(t+\tau|{\bf \Gamma},t)
-
T\int_0^{\tau} \mathrm{d}t'\;
 \nabla_{p,j\alpha} S^\mathrm{perm}({\bf p}(t+t'))
\nonumber \\ & = &
{q}_{j\alpha}^{0,\mathrm{cl}}(t+\tau|{\bf \Gamma},t)
-s_{j\alpha} k_\mathrm{B} T
\int_0^{\tau} \mathrm{d}t'\;
 \ln \left[\frac{ N_{{\bf a}_j'}({\bf p}(t)) +1
 }{N_{{\bf a}_j}({\bf p}(t)) } \right]
\nonumber \\ &  & \mbox{ } \times
\delta\big( p_{j\alpha}- (a_{j\alpha}+a'_{j\alpha})/2\big)
\nonumber \\ & = &
{q}_{j\alpha}^{0,\mathrm{cl}}(t+\tau|{\bf \Gamma},t)
 -\frac{s_{j\alpha} k_\mathrm{B} T}{|\dot p_{j\alpha}^{0,\mathrm{cl}}(t)| }
\ln \frac{ N_{{\bf a}_j'}({\bf p}(t)) +1 }{N_{{\bf a}_j}({\bf p}(t)) }
\nonumber \\ & = &
{q}_{j\alpha}^{0,\mathrm{cl}}(t+\tau)
 -
 \frac{ k_\mathrm{B} T}{\dot {p}_{j\alpha}^{0,\mathrm{cl}}(t) }
\ln \frac{ N_{{\bf a}_j'}({\bf p}(t)) +1 }{N_{{\bf a}_j}({\bf p}(t)) } ,
\;\;
a_{j\alpha}' \ne a_{j\alpha} .\nonumber
\end{eqnarray}
This uses the usual rule for evaluating the integral
of a Dirac-$\delta$ function of a function.
Note that the classical adiabatic rate of change of momentum is just
the classical force,
$\dot {{\bf p}}^{0,\mathrm{cl}} = {\bf f}^0$.
This result for the gradient of the permutation entropy
can also be used to evaluate the dissipative force.

The second term on the right hand side
of the final equality
is independent of the size of the time step $\tau$.
It can be large whenever a boson changes momentum states
with markedly difference occupancies,
or whenever the classical force on it is small,
which may be computationally problematic.

In the classical regime most states are empty
and the occupied states contain at most one boson,
and so $N_{{\bf a}_j}({\bf p}(t)) = 1$
and $N_{{\bf a}_j'}({\bf p}(t)) = 0$.
In this case
the second term on the right hand side is zero.
This says that there are no velocity spikes or jumps in position
when an uncondensed  boson changes momentum state.

One can gain some insight into this expression as follows.
By design, the decoherent quantum equations of motion conserve entropy
on a trajectory.
Hence for a time step one should have that $\Delta S = 0 $,
or
\begin{eqnarray}
\Delta S^\mathrm{perm} & = &
-\Delta S^\mathrm{r}
\nonumber \\ & = &
\frac{1}{T} \left[ \Delta {\cal K} + \Delta U \right]
\nonumber \\ & = &
\frac{1}{T} \left[ \frac{\tau {\bf p}\cdot{\bf f}^0}{m}
+
U({\bf q}^{0}(t+\tau)) - U({\bf q}(t)) \right]
\nonumber \\ & \approx &
\frac{1}{T} \left[ \frac{\tau {\bf p}\cdot{\bf f}^0}{m}
-
[{\bf q}^{0}(t+\tau) - {\bf q}(t)] \cdot {\bf f}^0  \right].
\end{eqnarray}
Inserting the above result for the decoherent quantum change in position gives
\begin{eqnarray}
\Delta S^\mathrm{perm} & = &
\frac{1}{T} \left[ \frac{\tau {\bf p}\cdot{\bf f}^0}{m}
-
\frac{\tau {\bf p}\cdot{\bf f}^0}{m}
\right. \nonumber \\ && \mbox{ } \left.
+
\sum_{j,\alpha} f_{j\alpha}^0
 \frac{ k_\mathrm{B} T}{\dot {p}_{j\alpha}^{0,\mathrm{cl}}(t) }
\ln \frac{ N_{{\bf a}_j'}({\bf p}(t)) +1 }{N_{{\bf a}_j}({\bf p}(t)) }
 \right]
\nonumber \\ & = &
k_\mathrm{B}
\sum_{j}\!^{({\bf a}_{j}' \ne {\bf a}_{j})}
\ln \frac{ N_{{\bf a}'_j}({\bf p}(t)) +1 }{N_{{\bf a}_j}({\bf p}(t)) }
\nonumber \\ & = &
k_\mathrm{B}
\sum_{{\bf a}}
\ln \frac{ N_{{\bf a}}({\bf p}(t+\tau))! }{N_{{\bf a}}({\bf p}(t))! } .
\end{eqnarray}
The final step assumes that a change of momentum state is a rare event:
for boson $j$ changing state,
${\bf a}_{j} \Rightarrow {\bf a}_{j}'$,
the result
$N_{{\bf a}_{j}}(t+\tau) = N_{{\bf a}_{j}}(t)-1$
and
$N_{{\bf a}_{j}'}(t+\tau) = N_{{\bf a}_{j}'}(t)+1$
only holds if no other changes to these states occur in this time step.
In formal analysis one can always choose the time interval $\tau$ small enough
so that there is at most one change in momentum state during it,
and that this only occurs in one component at a time.
In general
the right hand side of the final equality
is the definition of the change in permutation entropy.

There are therefore two challenges
in implementing this expression
for the gradient of the permutation entropy
in the discrete case.
First the assumption that changes in momentum state are rare
and occur one boson at a time
 might necessitate an unfeasibly small time step.
And second, as the fourth equality shows,
it relies upon a small change in position.
Hence the first order equations of motion break down
when the contribution from the gradient of the permutation entropy is large,
as occurs  for the components in which
the disparity in occupation between the states is large,
or in which the classical force is small.

%
\section{Ansatz for Constant Entropy Trajectory} \label{Sec:ConstEntTraj}
\setcounter{equation}{0} \setcounter{subsubsection}{0}
\renewcommand{\theequation}{\Alph{section}.\arabic{equation}}
%

Here  is offered an alternative algorithm
that maintains the total entropy constant
on the decoherent quantum trajectory,
but without the need for continuous occupancy
and umbrella sampling correction.

With ${\bf \Gamma}_1 = {\bf \Gamma}(t_1)$,
let $\tilde {\bf \Gamma}_2 = {\bf \Gamma}^\mathrm{0,cl}(t_2|{\bf \Gamma}_1,t_1)$,
be the classical adiabatic evolution at the next time node, $t_2 = t_1+\tau$.
Let $\Delta {\bf q}(t_2) = {\bf q}(t_2) - \tilde{\bf q}(t_2)$
be the shift to the actual position at $t_2$ due to the permutation entropy,
an expression for which is now obtained.

The change in permutation entropy due to the classical adiabatic evolution is
$ \Delta S^\mathrm{perm} =
S^\mathrm{perm}(\tilde{\bf p}(t_2)) - S^\mathrm{perm}({\bf p}(t_1))$.
The change in permutation entropy must cancel with the change in reservoir entropy
due to the shift,
$S^\mathrm{r}({\bf q}(t_2)) - S^\mathrm{r}(\tilde{\bf q}(t_2))
= -\Delta S^\mathrm{perm}$,
or
\begin{eqnarray}
\Delta S^\mathrm{perm}
& = &
\frac{1}{T} [U({\bf q}(t_2)) - U(\tilde{\bf q}(t_2))]
\nonumber \\ & \approx &
\frac{-1}{T} \Delta {\bf q}(t_2) \cdot {\bf f}^0(\tilde{\bf q}(t_2)) .
\end{eqnarray}
The final approximation is valid when the change in position is small
enough that a linear expansion of the potential is justified.
Contrariwise,
divide the interval into $n+1$ nodes,
$\{ {\bf q}^{(0)}, {\bf q}^{(1)}, \ldots, {\bf q}^{(n)} \}$,
with ${\bf q}^{(0)}=\tilde{\bf q}(t_2)$
and ${\bf q}^{(n)}={\bf q}(t_2)$,
such that
\begin{equation}
\frac{1}{n}\Delta S^\mathrm{perm}
=
\frac{-1}{T} [{\bf q}^{(j+1)}-{\bf q}^{(j)}]
\cdot  {\bf f}^0({\bf q}^{(j)}) ,
\;\; j = 0,1,\ldots,n-1 .
\end{equation}
This gives ${\bf q}(t_2)$ by stepping successively along the path
with the ansatz
\begin{equation}
{\bf q}^{(j+1)} = {\bf q}^{(j)} + c^{(j+1)} {\bf f}^0({\bf q}^{(j)}),
\;\; j = 0,1,\ldots,n-1 .
\end{equation}
This minimizes the root mean square departure from the classical path.
The scale factor is given by
\begin{equation}
c^{(j+1)}
=
\frac{-T\Delta S^\mathrm{perm} /n
}{ {\bf f}^0({\bf q}^{(j)}) \cdot  {\bf f}^0({\bf q}^{(j)}) } ,
\;\; j = 0,1,\ldots,n-1 .
\end{equation}

As a result of this the gradient of the permutation entropy
that appears in the equations of motion to keep the total entropy constant is
\begin{equation}
\nabla_p S^\mathrm{perm}({\bf p}(t_1))
=
\frac{- 1}{\tau T} \Delta {\bf q}(t_2) ,
\end{equation}
with the right hand side now known.
Accordingly,
the dissipative force for the thermostat is
\begin{eqnarray}
\overline {\bf R}_p(t_1)
& = &
\frac{\sigma^2}{2k_\mathrm{B}}
\nabla_p \left[ S^\mathrm{r}({\bf \Gamma}(t_1))
+  S^\mathrm{perm}({\bf \Gamma}(t_1)) \right]
\nonumber \\ & = &
\frac{\sigma^2}{2k_\mathrm{B}}
\left[ \frac{-1}{mT}{\bf p}(t_1)
- \frac{1}{\tau T} \Delta {\bf q}(t_2)  \right].
\end{eqnarray}
The  change in momentum due to the thermostat
should be added to the classical adiabatic change,
${\bf p}(t_2) =
\tilde {\bf p}(t_2)
+ \overline {\bf R}_p(t_1) + \tilde {\bf R}_p(t_1)$.

The computational challenge with the proposed algorithm
is that it increases the cost of a time step
by about a factor of $n$,
since this is the number of times the force needs to be calculated.
With the neighbor table, the cost of evaluating the force
is linear in $N$.
There are possible savings on these estimates because the current
second order algorithm already evaluates the gradient of the force.

One should be cautious in considering this proposed ansatz.
The trajectories in subsection~\ref{Sec:Traj}
show that the jump in position
corresponds to a boson approaching the boundary of its momentum state,
and it is parallel or anti-parallel to that component of the corresponding force.
The jumps sometimes allow and sometimes preclude the momentum state transition.

\vfill

\end{document}